%% file: twisted_graphite.tex
\documentclass[onecolumn,amssymb, nobibnotes, floatfix, aps, prl]{revtex4-2}

\usepackage{hyperref}
\usepackage[utf8]{inputenc}
\usepackage[english]{babel}
\usepackage{amssymb}
\usepackage{amsmath}
\usepackage{graphicx}
\usepackage{color}
\usepackage{calc}
\renewcommand{\vec}[1]{\boldsymbol{\mathrm{#1}}}

\begin{document}

\title{Twists and The Electronic Structure of Graphitic Materials} 

\author{Tommaso Cea$^1$}
\email{tommaso.cea@imdea.org}
\author{Niels R. Walet$^2$}\email{Niels.Walet@manchester.ac.uk}
\author{Francisco Guinea$^{1,3}$}
\email{paco.guinea@imdea.org}
\affiliation{$^1$Imdea Nanoscience, Faraday 9, 28015 Madrid, Spain}
\affiliation{$^2$Department of Physics, University of Manchester, Manchester, M13 9PY, UK}
\affiliation{$^3$School of Physics and Astronomy, University of Manchester, Manchester, M13 9PY, UK}

\date{\today}

\begin{abstract}
We analyze the effect of twists on the electronic structure of configurations of infinite stacks of graphene layers. We focus on three different cases: an infinite stack where each layer is rotated with respect to the previous one by a fixed angle, two pieces of semi-infinite graphite  rotated with respect to each other, and finally a single layer of graphene rotated with respect to a graphite surface. In all three cases we find a rich structure, with sharp resonances and flat bands for  small twist angles. The method used can be easily generalized to more complex arrangements and stacking sequences.
\end{abstract}

\maketitle

\section{Introduction}
The discovery of both superconductivity and insulating behavior in twisted graphene bilayers\cite{Ketal17,Cetal18a,Cetal18b}, see also\cite{Hetal18,Yetal19}, has lead to an extraordinary research effort on twisted graphene systems. The twist between neighboring layers leads to the formation of a Moir\'e superlattice. The size of the unit cell increases as the twist angle is reduced. For sufficiently large lattice units, we can ignore the atomistic nature of the carbon atoms, and describe the low-energy electronic bands by a continuum model \cite{LPN07,M10,BM11,M11,LPN12}. These superlattice bands are very narrow, with a bandwidth of a few meV's for certain twist angles\cite{SCVPB10,TMM10,BM11}. This analysis has been generalized to twisted trilayers\cite{AC18,MRB19}, tetralayers\cite{ZMCJS19,CCJ19}, and to some twisted infinite stacks\cite{KKTV19}.

The problem of twists is related to the occurrence of 
stacking defects in graphite and at graphite surfaces. Both of these have been extensively studied experimentally, see Refs.~\cite{KCS90,WRGBB09,FCCV13,Yetal14}, and Refs.~\cite{Metal05,LA07,LLA09,Netal09,Zetal18}. The electronic properties of stacking defects in graphite have also been studied theoretically\cite{GNP06,AG08}. It is worth noting that single graphene layers on graphite can be manipulated with atomic scale probes\cite{Xetal12}, and that unusual superconducting like features in bulk graphite have been ascribed to stacking misalignment\cite{EHLTV14,BEB15}. 

\begin{figure}
\includegraphics[width=\columnwidth]{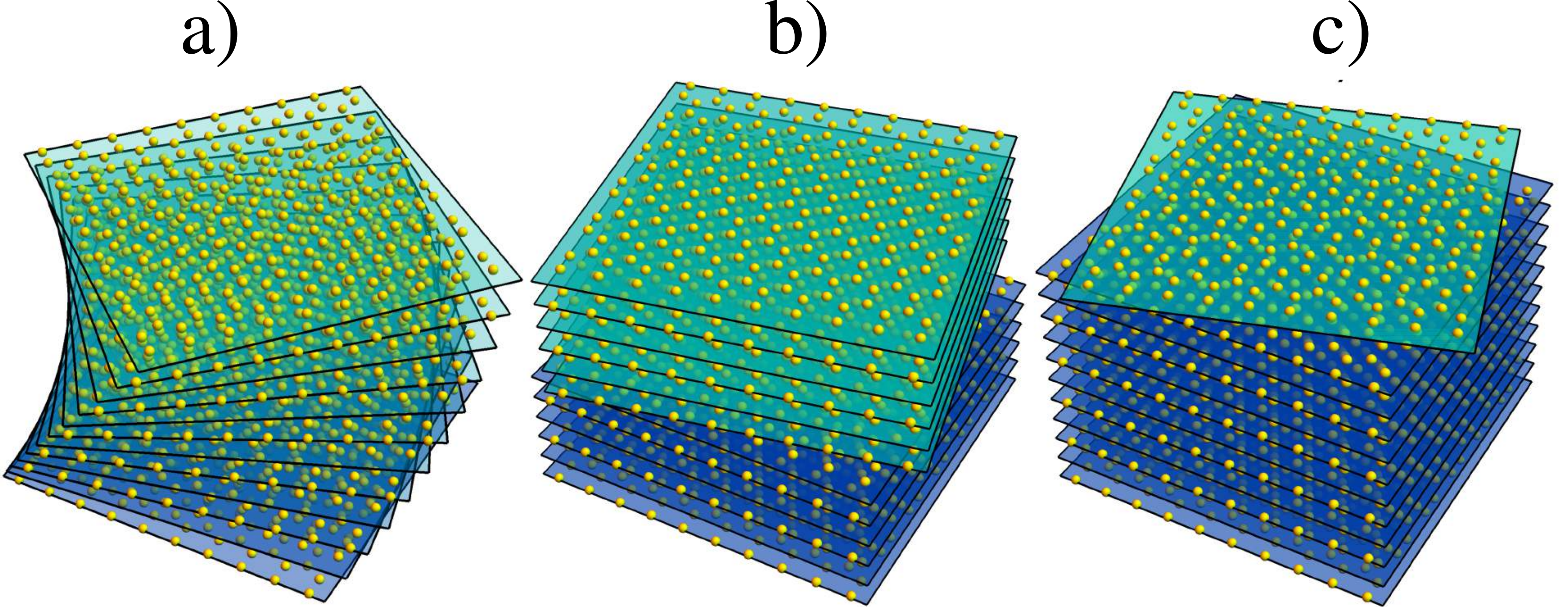}
\caption{Schematic representation of the structures studied: a) Infinite stack of graphene layers, where each successive layers is rotated by an angle $\theta$ relative to the previous one. b) Stacking fault where two semi-infinite graphite crystals are rotated by an angle $\theta$. c) Twisted graphene layer on a graphite surface.}
\label{fig:sketch}
\end{figure}

In the following, we study three representative cases of three dimensional systems with a twist.
A sketch of these structures is shown in Fig.~\ref{fig:sketch}. As described below, the analysis presented here can easily be extended to other intermediate situations. 
The first case is  a stack of a graphene layers where each layer is rotated by a constant amount with respect to the previous one. This arrangement can be achieved by applying a global twist to three-dimensional graphite, provided that neighboring layers slide and lose registry at the atomic scale in the same way. This structure is a natural generalization to three dimensions of a twisted graphene bilayer.
The two other situations both describe rotations involving stacks of bulk graphite. We use the Bernal stacking sequence, $ABAB , \cdots$ for graphite, although extension to other sequences is straightforward. In the second case we consider two semi-infinite graphite crystals twisted with respect to each other, and the final one is  a single graphene layer twisted with respect to a graphite surface.  
The twist between layers leads to narrow bands in many cases, which will enhance the effect of the electron-electron interactions and may lead to broken-symmetry phases. The study of such features lies beyond the scope of this work. 

We now turn to a detailed description of the three models.
 In the continuum model, we take the parametrization of the interlayer hoppings between the twisted layers from Ref.~\cite{KYKOKF18}, where the  $AA$ and $AB$ hoppings are described by the two parameters $\{ g_1 , g_2 \} = \{ 79.7 , 97.5 \}\,\mathrm{meV}$.
 We will use a twist angle $\theta = 1.08^\circ$, which is the first magic angle for the parametrization given above.

\section{Infinite Stack of Twisted Layers}
In twisted bilayer graphene we find a Moir\'e superlattice, which does not require perfect alignment. A
a commensurate superlattice at the atomic scale can only be defined for certain angles where reciprocal vectors of the two layers coincide. The continuum approximation developed in Refs.~\cite{LPN07,BM11} describe an approximate Moir\'e superlattice for any twist angle, with a lattice unit $L = d / ( 2 \sin ( \theta / 2 )$, where $d$ is the lattice unit of graphene and $\theta$ is the twist angle. This continuum description has been shown to provide an excellent approximation to the electronic bands at commensurate angles at the atomic scale, and is expected to describe the properties of the bilayer at arbitrary angles. 

The general analysis of the electronic properties for more than two misaligned layers is more challenging, as the superposition of different Moir\'e patterns leads  to complex structures with no obvious periodicity. A general analysis for the case of three rotated layers is given in Ref.~\cite{AC18}. The complexity of that method increases exponentially with the number of layers. The situation simplifies when the twist angle between successive layers is of equal magnitude and opposite sign between neighboring pairs of layers \cite{KKTV19}. 
In the case of constant twist angle, even in the simple case of a trilayer, the Moir\'e structures defined by each pair of layers, which are of equal size, do not coincide, and a full description requires the consideration of many additional reciprocal vectors. As the angle is reduced, the misalignment between the two Moir\'e patterns decreases. By assuming that the two interlayer rotations define exactly the same Moir\'e pattern, a continuum approximation to the bands of the twisted trilayer can still be obtained\cite{MRB19}, where the required number of reciprocal lattice vectors is of the same order of magnitude as that for a twisted bilayer.

We describe the infinite rotated stack as a succession of trilayers embedded in an environment to be defined self consistently, in an analogous manner  as the Coherent Phase Approximation\cite{S67}. We use three sets of in-plane reciprocal lattice momenta, one for each layer, labeled by a two-dimensional vector, $\vec{ k}_\parallel^i , i = 1, 2, 3$. These are then repeated periodically in the $z$ direction. In a similar way as for the bilayer \cite{BM11}, each set of momenta is defined in a triangular lattice with reciprocal lattice spacing $| \vec{ G} | = (4 \pi ) / ( \sqrt{3} L )$.
We introduce a wavevector $k_z$ along the $z$ direction. The relative phase between successive trilayers is $k_z \bar{d}_z$, where $\bar{d}_z$ is three times the interlayer distance, $c$. This allows us to classify the eigenstates as function of $\vec{k}_\parallel$ and $k_z$, where $\vec{k}_\parallel$ is the same for the three layers.
We consider only couplings between nearest-neighbor layers. Couplings between second-nearest-neighbor layers, given in the Bernal stacking by the hopping parameter $\gamma_2$, (see below) give rise to couplings between $AA$ sites modulated over the  Moir\'e supercell. We have not included these terms, since they are only a small perturbation to the hamiltonian used here. 

The assumption of a well defined Moir\'e structure is only valid for small angles. As the angle increases, the misalignment between layers at a given distance also increases. This misalignment tends to decouple the layers, and thus the model used here overestimates the interactions between distant layers, so that our results may overestimate the dispersion along the $k_z$ direction.
 Our framework assumes that each block of three consecutive twisted layers defines an approximate Moir\'e pattern. The same approach applies for different stacking arrangements: $AAA$, $ABA$ or  $ABC$. If the twist angle is small enough, the approximate Moir\'e pattern identified by each block varies slowly along the stack. As we assume that we need to consider only nearest-neighbor interlayer couplings, we can locally describe the stack as a single Moir\'e pattern, disregarding the type of stacking arrangement.

\begin{figure}
\includegraphics[width=\columnwidth]{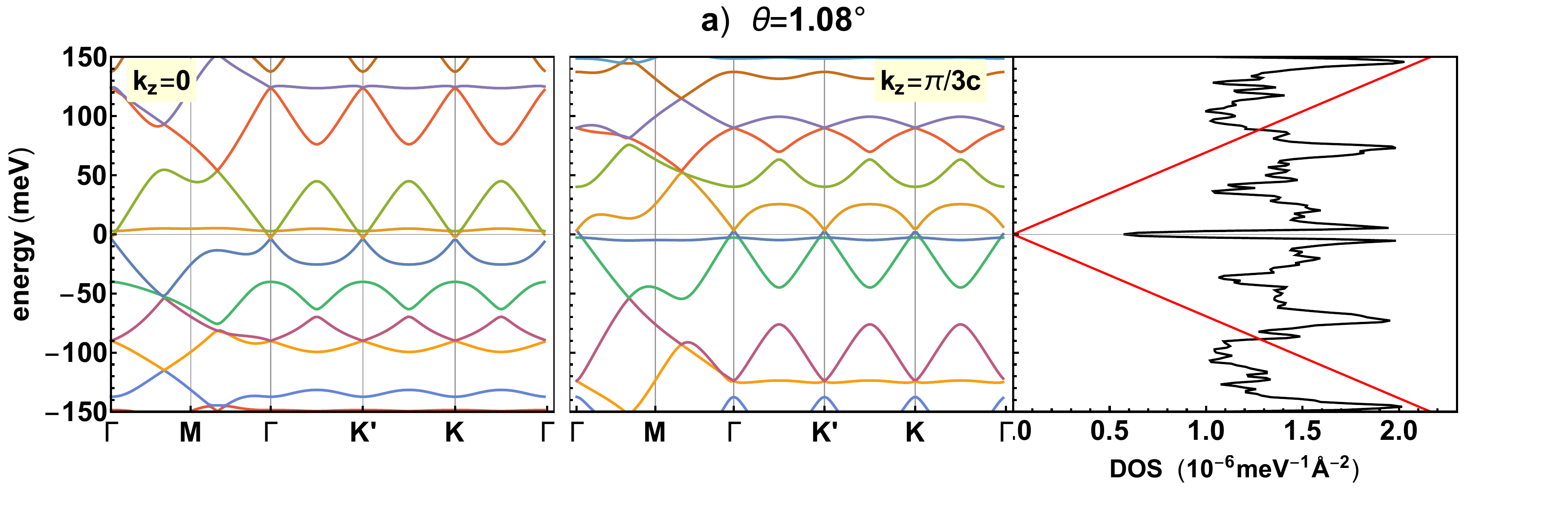} 
\\
\includegraphics[width=\columnwidth]{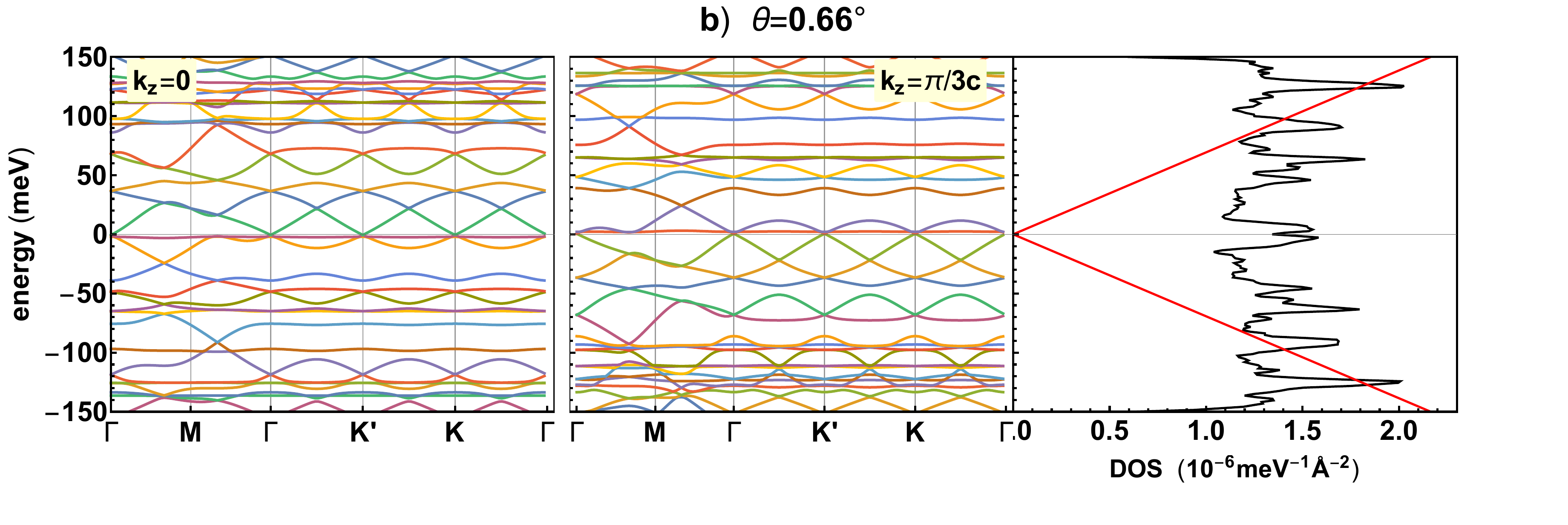} 
\\
\includegraphics[width=1.5in]{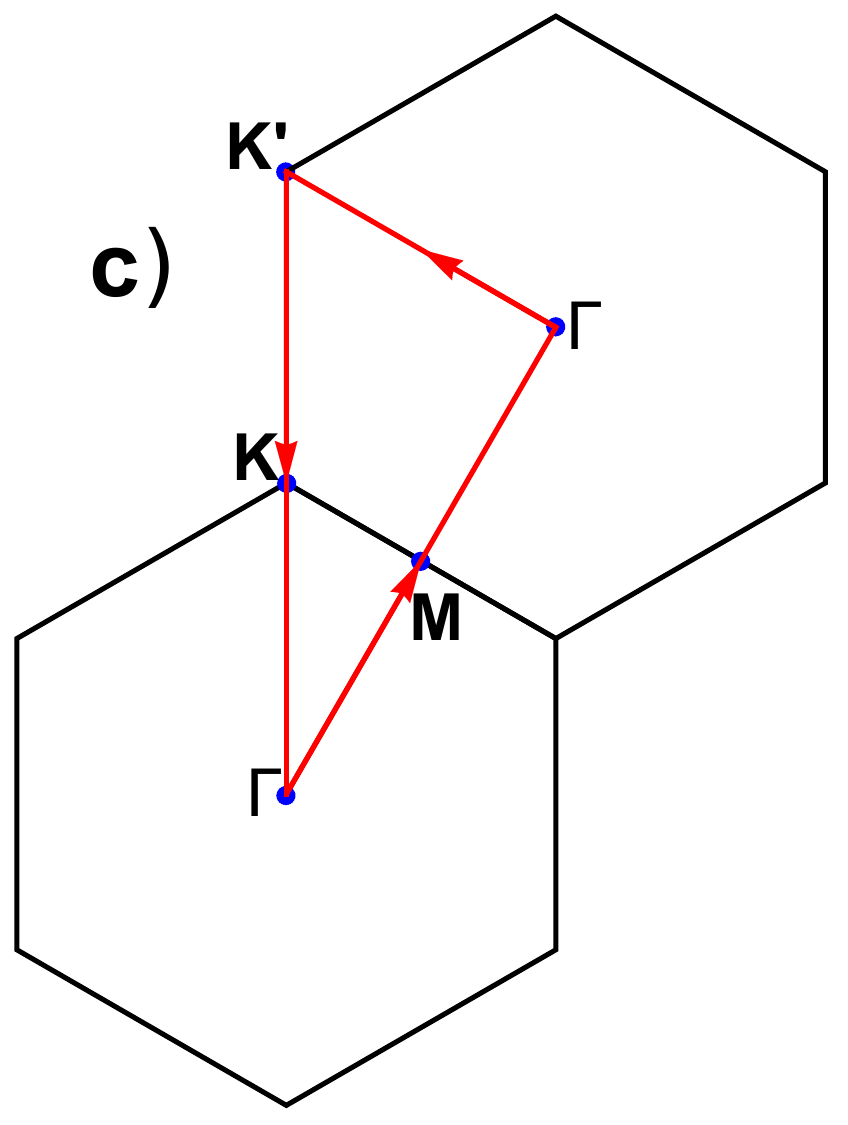}
\hspace{0.3cm}
\includegraphics[width=2.5in]{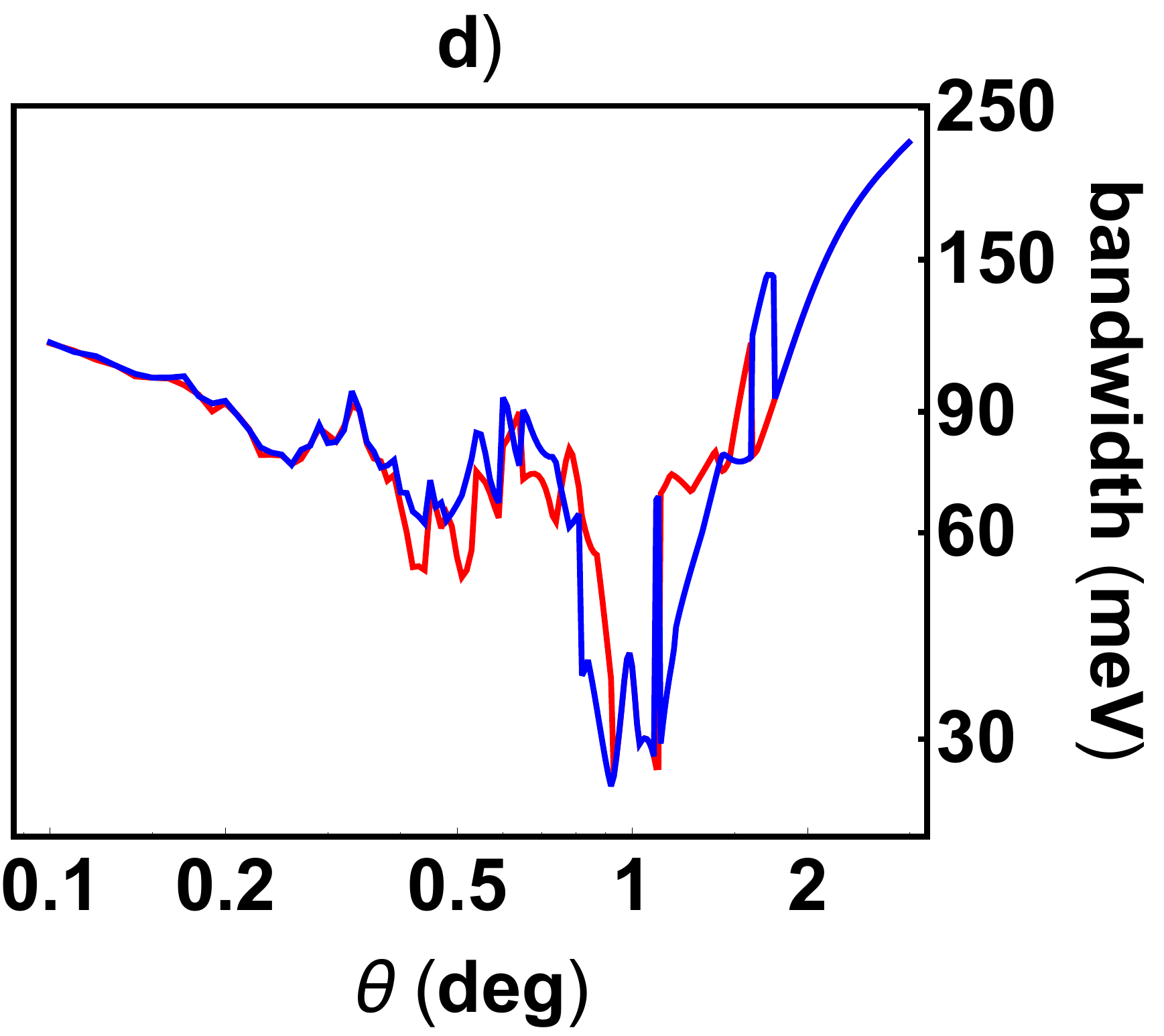} 
\caption{a),b): electronic bands and corresponding densitiy of states (DOS) for an infinite stack of  twisted graphene layers for two different twist angles between neighboring layers. The bands are labelled by the momentum parallel to the layers, $\vec{ k}_\parallel$, and the momentum perpendicular to the layers, $k_z$. The unit cell contains three layers. The red line superimposed to the densities of states represents the case of three uncoupled layers.
c): schematic of the path in the reciprocal space along which the bands are computed.
d): bandwidth of the two bands closest to the neutrality point, computed along the high symmetry path, as a function of the angle using a logarithmic scale.
}
\label{fig:bands_twisted}
\end{figure}

Typical bands, as function of $\vec{ k}_\parallel$ and $k_z$  are shown in  Fig.~\ref{fig:bands_twisted}a,b, along with the corresponding density of states.

The results show the coexistence of very narrow and wider bands over a broad range of energies. The bands at $k_z = 0$ and $k_z = \pi$ are equivalent under the simultaneous transformations $\vec{ k}_\parallel \leftrightarrow - \vec{ k}_\parallel$ and $E \leftrightarrow - E$. The narrow bands give rise to sharp peaks in the density of states. The overlapping of narrow and broad bands is reminiscent of a twisted trilayer\cite{MRB19,si}.
We find two sharp minima in the width of the lowest energy bands for $\theta\simeq1^\circ$ (see Fig.~\ref{fig:bands_twisted}d)), suggesting the possibility of defining two ``magic angles''. The charge density distribution within the unit cell as function of wavevector is shown in \cite{si}. The charge densities for the narrowest band near the neutrality point are very similar for the $K, K'$ and $\Gamma$ points. This is rather different from the significant momentum dependence of the charge distributions in twisted graphene bilayers\cite{RM18,GuineaWalet18}.

\section{Twisted Stacking Fault in Bulk Graphite}
We now consider two semi-infinite graphite stacks rotated with respect to each other by an angle $\theta$. The misalignment at the interface between the two stacks induces a Moir\'e pattern, in an analogous case to a twisted graphene bilayer. The standard calculation of the band structure of a twisted graphene layer\cite{LPN07,BM11} requires the knowledge of the Green's function of the states with momenta mixed by the Moir\'e superstructure. For the twisted bilayer this Green's function can be obtained from the inversion of the hamiltonian of each individual layer, $G_i ( \vec{ k}_\parallel , \omega ) = ( \omega {\cal I}_{\sigma_i} - {\cal H}_{i , \vec{ k}_\parallel} )^{-1}$, where $\omega$ is the frequency, $i = 1 , 2$ is a layer index, ${\cal I}_\sigma$ is a $2 \times 2$ matrix defined in the sublattice space of layer $i$, and ${\cal H}_{i , \vec{ k}_\parallel}$ is a $2 \times 2$ matrix which defines the hamiltonian in layer $i$. In the case of a defect embedded in a three dimensional stack, the Green's function at each side of the defect can be expressed in terms iof the slef energy, $G_i ( \vec{ k}_\parallel , \omega ) = [ \omega {\cal I}_{\sigma_i} - \Sigma_i ( \omega , \vec{ k}_\parallel ) ]^{-1}$, where $\Sigma_i ( \omega , \vec{ k}_\parallel )$ has to be chosen so that $G_i ( \vec{ k}_\parallel , \omega )$ gives the Green's function at the surface of a semi-infinite stack. These self energies can be obtained by recursive methods\cite{HHK72,GTFL83,si}. The required interlayer hopping parameters that describe the bands of bulk graphite are well known in terms of the SWM parameters, and  have been extensively studied \cite{M57,SW58,DD02,NGPNG09}.

\begin{figure}
\includegraphics[width=\columnwidth]{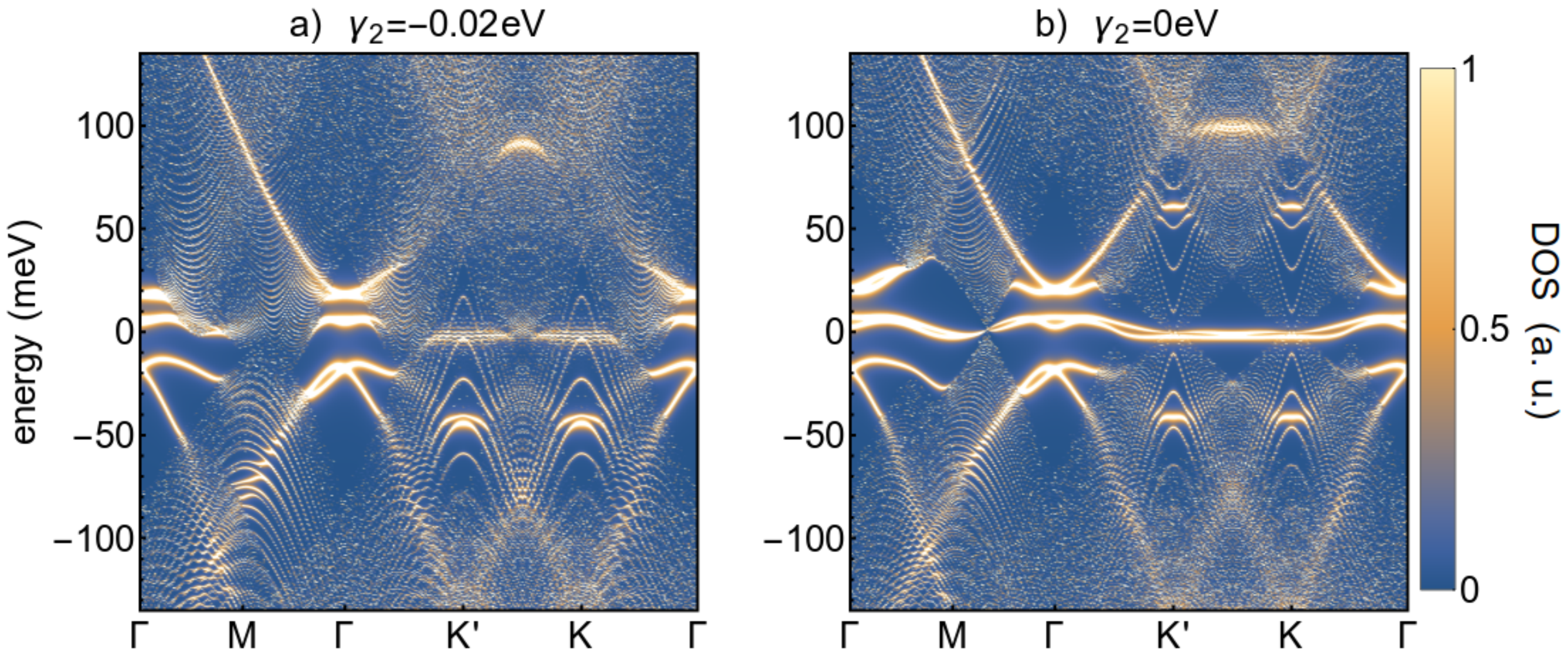} 
\caption{Two-dimensional electronic density of states 
at  a twist defect between two semi-infinite graphite stacks, with twist angle $\theta = 1.08^\circ$. We show the influence of the hopping  $\gamma_2$ between second-nearest-neighbor layers.}
\label{fig:bands_twisted_stacks}
\end{figure}

Results for the density of states of the two layers at the stacking defect are shown in Fig.~\ref{fig:bands_twisted_stacks}. The nature of the bands, or resonances, at very low energies depends sensitively on the value of the hopping between second-nearest-neighbors layers, $\gamma_2$, and results for the graphite value $\gamma_2$ and $\gamma_2=0$ are shown. This parameter is the smallest and the least precisely determined of the set which describes the electronic structure of bulk graphite. 

The assumption $\gamma_2=0$ in modelling graphite leads to a nodal line of Fermi points instead of the observed two-dimensional Fermi surface. A finite value of $\gamma_2$ is also needed in order to explain the finite bandwidth of the $n=0$ Landau level of graphite\cite{OS66,Aetal17,Jetal19}. Experiments in bulk graphite are consistent with a value $\gamma_2 \approx -0.02\,\text{eV}$, see, for instance\cite{Setal09,Setal12}. On the other hand, recent measurements on 4-8 Bernal-stacked multilayers \cite{NKKMM16,Getal16,KSKM17,NKSM18} suggest that the second-nearest-neighbor interlayer hopping is much smaller, $\gamma_2 \sim 0$.

The broadening of the features shown in Fig.~\ref{fig:bands_twisted_stacks} is due to the overlap of states localized at the defect with continuum states in bulk graphite. A simple analysis of the bulk bands projected onto a two-dimensional surface shows that the edges of the continuum lie approximately at $\epsilon ( \vec{ k}_\parallel) = \pm \gamma_2 \pm ( 3 \gamma_0  | \vec{ k}_\parallel | d ) |^2)/(8 \gamma_1) \pm (\sqrt{3} \gamma_3 | \vec{ k}_\parallel | d)/2$, 
where we use the conventional notation for the band structure of graphite \cite{DD02}, $\gamma_4$ and $\gamma_5$ have been neglected, and we assume that $\gamma_2 \ll \gamma_0 ,\gamma_1 , \gamma_3$. The above expression for $\epsilon ( \vec{ k}_\parallel)$ shows that states at $\epsilon = 0$ do not overlap with the bulk continuum except for $| \vec{ k}_\parallel | = 0$. For $\gamma_2 \ne 0$, and neglecting the effect of $\gamma_3$, the states at $\epsilon = 0$ overlap with the continuum for $| \vec{ k}_\parallel | = k_0 \lesssim \sqrt{8 \gamma_1 \gamma_2} / ( 3 \gamma_0 d )$. The distance between the $\Gamma$ and $K$ points in the Moir\'e Brillouin Zone is $k_{\Gamma-K} = [ 3 \sin ( \theta / 2 ) ] / ( 2 \pi d )$. For $\theta \lesssim 0.4^\circ$, we find $k_{\Gamma-K} \lesssim k_0$, and the entire Brillouin Zone overlaps with the continuum\cite{note2}. Note that the role of $\gamma_2$ in the case of the stacking fault considered here is very different from the role of $\gamma_2$ in the continuously twisted stack analyzed previously, as, in the present case, the series of second nearest neighbors at both sides of the defect are in registry.

\begin{figure}
\includegraphics[width=\columnwidth]{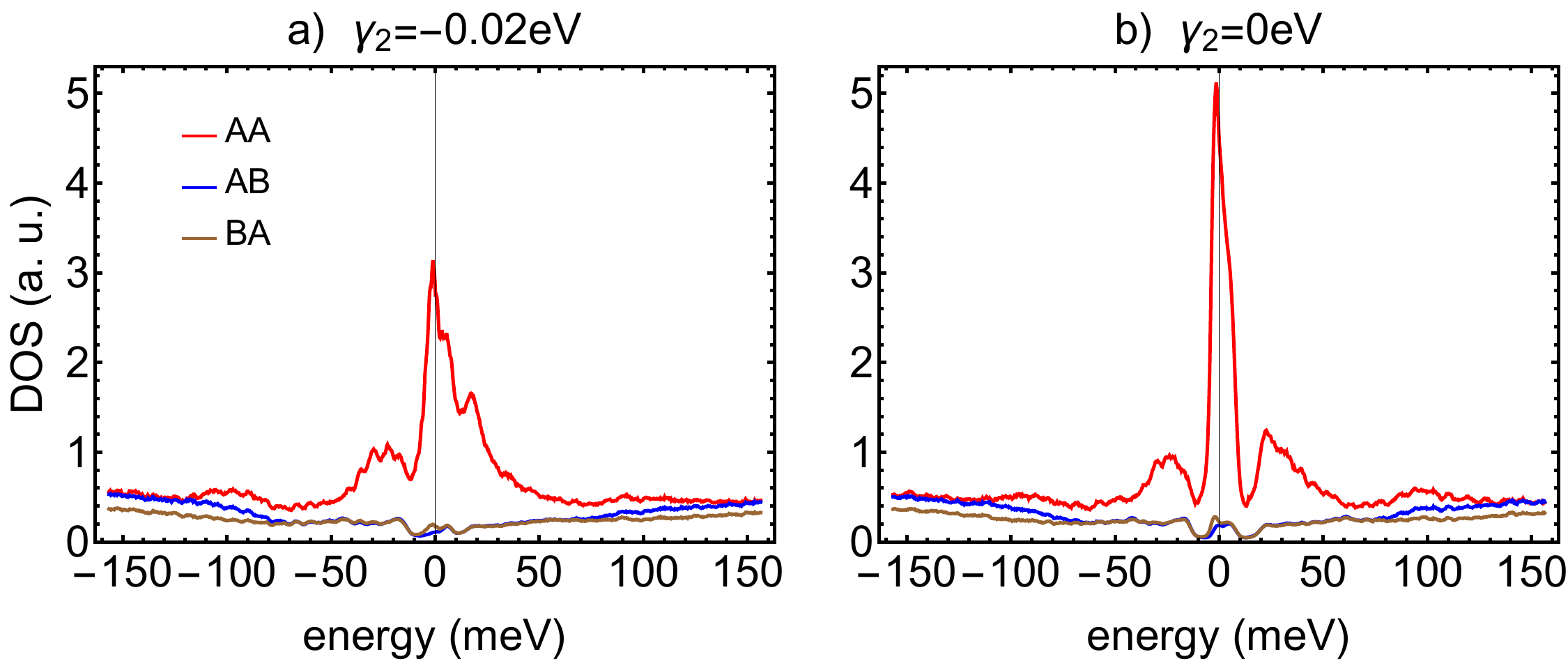}
\caption{
Local DOS on the boundary surface of the twisted stack shown in Fig.~\ref{fig:sketch}(b), computed in the$AA$ (red line), $AB$ (blue line) and $BA$ (brown line) aligned  regions of the Moir\'e unit cell, for $\gamma_2=-0.02\,\text{eV}$ (a) and $\gamma_2=0\,\text{eV}$ (b).}
\label{twisted_stack_local_DOS}
\end{figure}

Information about the real-space origin of the spectral features of the twisted stack shown in  Fig.~\ref{fig:sketch}(b) can be obtained from Fig.~\ref{twisted_stack_local_DOS}. Here we 
show the local DOS on the boundary surface, computed in the $AA$ (red line), $AB$ (blue line) and $BA$ (brown line) regions of the Moir\'e unit cell.
As can clearly be seen, the DOS in the $AA$ region strongly exceeds those in the $AB/BA$ regions at the Van Hove singularity, which is in good agreement with the scanning-tunneling-microscopy maps reported in twisted graphene bilayer \cite{jiang_nat19,Cetal19,Ketal19}.

\section{Twisted Layer on a Graphite Surface}
\begin{figure}
\includegraphics[width=\columnwidth]{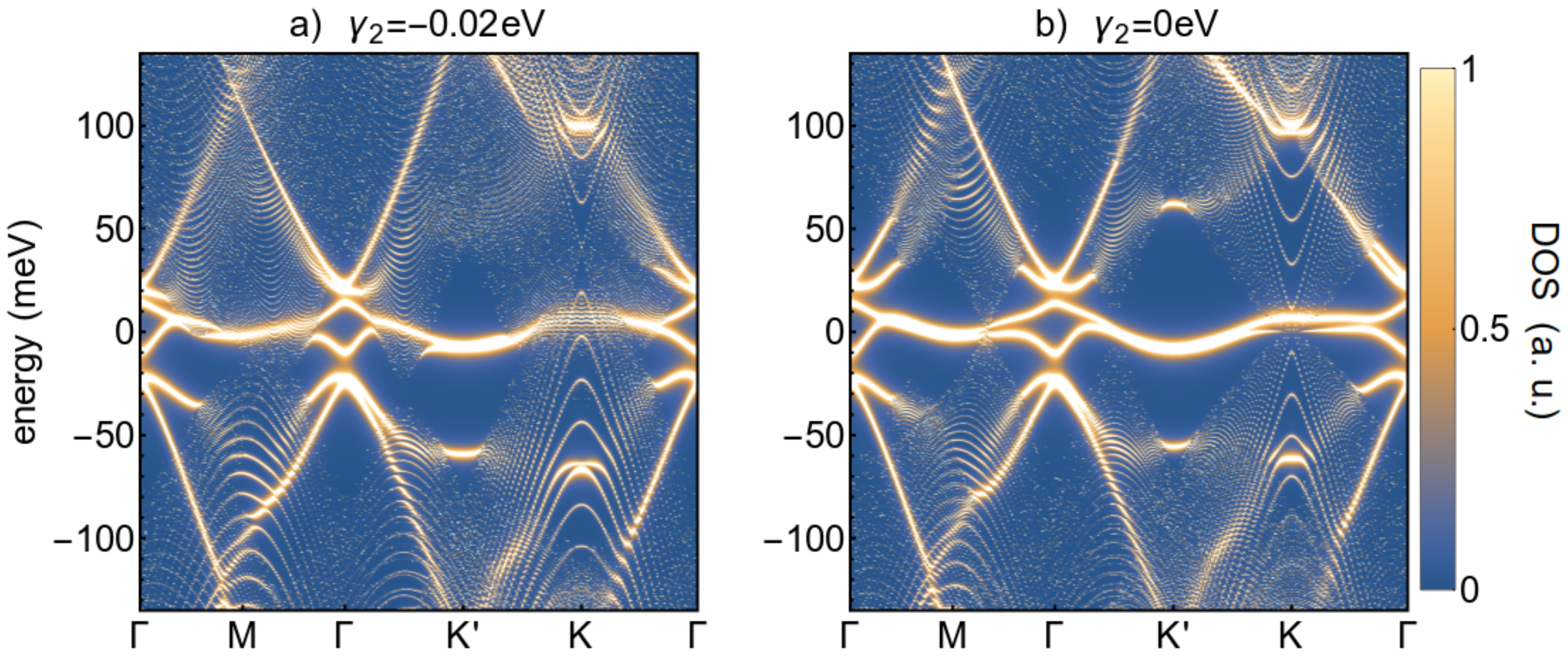}
\caption{Band structure for a single graphene layer rotated with respect to a graphite surface. See Fig.~\ref{fig:bands_twisted_stacks} for explanation.}
\label{fig:bands_surface_stacks}
\end{figure}

The analysis presented above can also be applied when a graphene layer is rotated with respect to the surface of a semi-infinite stack of graphene layers. In this case, the problem can be reduced to a twisted bilayer where the self-energy term influences only one layer. Results are shown in Fig.~\ref{fig:bands_surface_stacks}.

The results in Fig.~\ref{fig:bands_surface_stacks} are rather similar to those in Fig.~\ref{fig:bands_twisted_stacks}, but the features of the flat bands at low energies are sharper. The reduction in the value of $\gamma_2$ observed in not too large multilayers\cite{NKKMM16,Getal16,KSKM17,NKSM18}, might be related to an increase in the interlayer distance. If that is the case, a similar effect can take place at surfaces, enhancing the sharpness of the low energy resonances.

\section{Lattice Relaxation}
\begin{figure}
\includegraphics[width=\columnwidth]{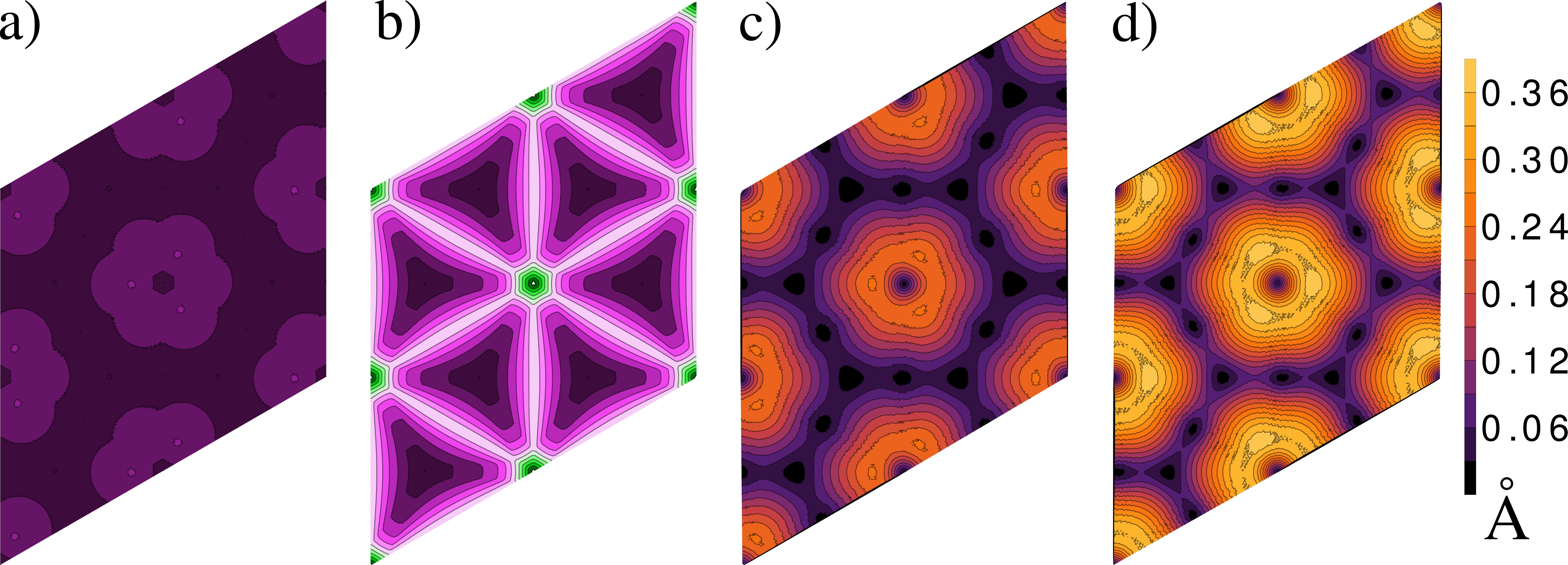} 
\caption{Lattice relaxation in a system made up of a graphene layer on a graphite surface. a),b) relative alignment between neighboring layers. Dark velvet corresponds to $AB$ and $BA$ alignment \cite{WG19,GW19}. Green stands for $AA$ alignment, and white gives the area of equal alignment (interface soliton). a) Shows the alignment between the top layer of graphite and the layer below, b) Shows the alignment between top layer of graphite and the twisted graphene layer.
c and d) absolute displacement of the atoms in the top graphite layer (c) and the graphene layer (d). A lighter color stands for large displacements. Black shows no displacement. The scale on the left labels the displacements are given in \AA.}
\label{fig:relaxation}
\end{figure}

Up to now we have assumed that there is no reconstruction of the lattice at the interface; it has been shown that this is an important effect for bilayers, see, e.g., Refs.~\cite{WG19,GW19,walet2019emergence}, and can have important consequences for the electronic structure as well. We focus on in plane relaxation, which is the largest contribution to the total relaxation \cite{nam_lattice_2017,WCFKT18,koshino_maximally-localized_2018,WG19,GW19}, especially for the (semi-)infinite structures studied here.
 We show results for  a graphene layer on a graphite surface, as, in this case, the top layer is weakly coupled to the rest of the system. We optimize the positions in the atoms in the twisted layer and in the topmost layer at the graphite surface. The intra-layer potential used is the AIREBO-M potential \cite{oconnor_airebo-m:_2015}, and the Kolomogorov-Crespi inter-layer potential \cite{kolmogorov_registry-dependent_2005}. We approximate the graphite as three layers of graphene, with the bottom two layers fixed, with one twisted graphene layer on top.
 Results are shown in Fig.~\ref{fig:relaxation}. Additional results for the relaxation of a twisted bilayer on a graphite surface are shown in \cite{si}. The plots in Fig.~\ref{fig:relaxation} show a significant relaxation of the topmost layer, in line with calculations for a twisted graphene bilayer\cite{WG19,GW19}. The first layer of the graphite stack is more weakly perturbed. The maximal displacement of atoms in the graphene layer  is $0.36\,\text{\AA}$, and in the top layer of the graphite $0.24\,\text{\AA}$. Comparing Fig.~\ref{fig:relaxation}b to  results for bilayer graphene in \cite{WG19,GW19}, we see that the deformation is less than that for a bilayer, but not by much. Since this would largely impact the hopping parameters at the interface, we expect that the electronic structure could be quite sensitive to this change.

\section{Conclusions}
We have analyzed the changes of the electronic properties due to small angle rotations between neighboring layers in three dimensional stacks of graphene layers. 

We first looked at a uniformly rotated stack, which corresponds to the application of a shear strain to bulk graphite. 
In this case we find both flat and dispersive bands, which overlap at low energies. We find, nevertheless, sharp resonances near the neutrality point at well defined twist angles.

In the case where one semi-infinite stack is rotated with respect to another, i.e., a stacking fault in bulk graphite, sharp features in the density of states appear in the defect region. Localized states with low dispersion exist within bulk gaps. The regions in energy and momenta where these gaps can be defined depend sensitively on the hopping between second-nearest-neighbor layers.

A single layer of graphene on a graphite surface shows similar low energy resonances and states within bulk gaps as a rotated stacking fault, although the resonances become sharper due to the disappearance of one half of the bulk "reservoir", and the coupling to this continuum  becomes weaker.

The momentum dependence of the charge density distributions within the Moir\'e unit cell varies in the three cases discussed here. In twisted graphite the charge is peaked at the $AA$ regions, independently of momentum. This is different from the case of twisted bilayer graphene, and suggests the existence of simple Wannier functions. The momentum dependence is significant for a graphene layer on a graphite surface. This dependence implies that the shape of the bands will be modified when the system is doped away from the neutrality point\cite{GuineaWalet18}, and an additional interaction, electron assisted hopping, can be defined. The strength of this interaction will be discussed elsewhere.

The results presented here will be sensitive to relaxation of the atomic positions at the interface, which we have also calculated. It is smaller than in bilayers, but not by a significant amount.

The problems discussed here are just one of many one could envisage in stacks of graphene layers. 
The method used for the last two problems relies on the substitution of the periodic stack of layers near the defect by an effective self energy.
This method is  very general, and it can be  extended to other combinations of Bernal, or rombohedral, graphite with rotated stacking faults and/or rotated layers at surfaces. Inhomogeneous electrostatic potentials induced by external gates can also be included, see~\cite{si}.\\\\

\acknowledgements
{\it Acknowledgements.}
We would like to thank Pablo San Jos\'e for useful conversations.
This work was supported by funding from the European Commission under the Graphene Flagship, contract CNECTICT-604391.

\input{twisted_graphite.bbl}

\pagebreak
\widetext
\begin{center}
\textbf{\large 
Supplementary Informations\\
Electronic Structure of Twisted Graphene Stacks,
Twisted Stacking Faults in Graphite, and Twisted Layers on Graphite Surfaces.
}
\end{center}
\setcounter{equation}{0}
\setcounter{figure}{0}
\setcounter{table}{0}
\setcounter{page}{1}
\makeatletter
\renewcommand{\theequation}{S\arabic{equation}}
\renewcommand{\thefigure}{S\arabic{figure}}
\renewcommand{\bibnumfmt}[1]{[S#1]}
\renewcommand{\citenumfont}[1]{S#1}

\newcommand{\be}{\begin{equation}}
\newcommand{\ee}{\end{equation}}
\newcommand{\bea}{\begin{eqnarray}}
\newcommand{\eea}{\end{eqnarray}}

\newcommand{\rmd}{{\rm d}}
\newcommand{\rmi}{{\rm i}}

\newcommand{\bra}[1]{\langle #1|}
\newcommand{\ket}[1]{|#1\rangle}
\newcommand{\braket}[2]{\langle #1|#2\rangle}

\renewcommand{\thefigure}{S\arabic{figure}}
\renewcommand{\thetable}{S\arabic{table}}
\renewcommand{\thesection}{S\arabic{section}}

\renewcommand{\theequation}{S\arabic{equation}}

\newcommand{\red}{\color{red}}
\newcommand\Niels[1]{\textcolor{cyan}{#1}}


\def\a{\alpha}
\def\b{\beta}
\def\e{\varepsilon}
\def\d{\delta}
\def\g{\gamma}
\def\m{\mu}
\def\l{\lambda}
\def\th{\theta}
\def\t{\tau}
\def\tt{\hat \tau}
\def\n{\nu}
\def\o{\omega}
\def\p{\pi}
\def\s{\sigma}
\def\phib{\bar \phi}
\def\phit{\tilde \phi}
\def\G{\Gamma}
\def\D{\Delta}
\def\O{\Omega}
\def\L{\Lambda}


\def\ra{\rightarrow}
\def\up{\uparrow}
\def\pll{\parallel}
\def\down{\downarrow}
\def\ran{\rangle}
\def\lan{\langle}
\def\Ra{\Rightarrow}
\def\pd{\partial}
\def\nb{\nabla}
\def\bk{{\bf k}}
\def\bl{{\bf l}}
\def\bq{{\bf q}}
\def\br{{\bf r}}
\def\bQ{{\bf Q}}
\def\bA{{\bf A}}
\def\bB{{\bf B}}
\def\bH{{\bf H}}
\def\bD{{\bf D}}
\def\bL{{\bf L}}
\def\bJ{{\bf J}}
\def\bM{{\bf M}}
\def\bR{{\bf R}}
\def\bS{{\bf S}}
\def\bn{{\bf n}}
\def\bx{{\bf x}}
\def\vecs{\vec{S}}
\def\nl{Nl$\s$m }
\def\LL{{\cal L}}
\def\DD{{\cal{D}}}
\def\SS{{\cal{S}}}
\def\OO{{\cal{O}}}
\def\cA{{\cal A}}

\def\nn{\nonumber}
\def\lb{\label}
\def\pref#1{(\ref{#1})}

\section{Infinite Stack of Twisted Layers. Continuum Model}
We consider the continuum model for an infinite stack of layers of graphene
with a constant small twist $\th$ between each layer and the consecutive one,
as depicted in the Fig. 1(a) of the main text.

 \begin{figure}[h]
\begin{tabular}{ccc}
\includegraphics[width=2in]{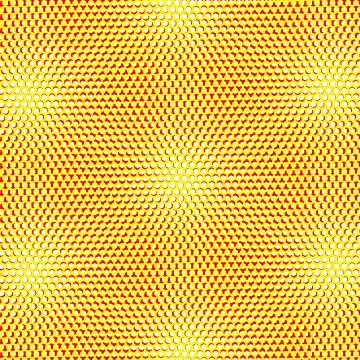} &
\includegraphics[width=2in]{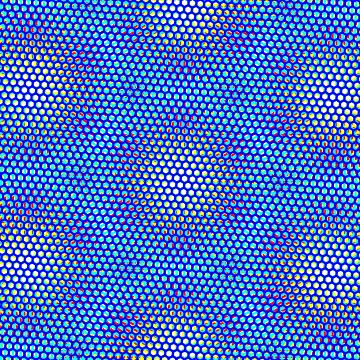} &
\includegraphics[width=2.in]{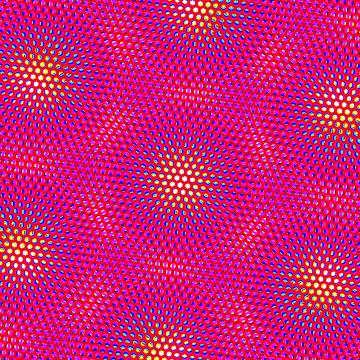} 
\end{tabular}
\caption{
Moir\'e patterns for two (left), four (center) and six (right) rotated honeycomb lattices.
The angle between successive layers is $\th=1.01^\circ$.
Only the left-hand case, with two layers, defines a commensurate Moir\'e pattern, with Moir\'e unit length $L\simeq56.7\times a$,
where $a$ is the unit length of each individual lattice.
An approximate Moir\'e pattern, with the same wavelength and orientation, is visible in the two other cases.}
\lb{moire_patterns}
\end{figure}
As mentioned in the main text,
a commensurate Moir\'e pattern cannot be defined in rotated graphene multilayers.
However, if the angle between successive layers is small and constant,
a relatively small number of layers still generates
an aprroximate periodicity similar to that found in
twisted bilayer for a general  twist angle where the layers are not commensurate.
An example is shown in Fig.~\ref{moire_patterns},
where we show the case of the approximate Moir\'e patterns formed by rotating two (left), four (center) and six (right) stacked layers.
If the out-of-plane hopping between layers is restricted to nearest neighbor ones,
one can attempt to describe the infinite structure in terms of the Moir\'e pattern identified by each pair of consecutive layers.
This argument can be made rigorous by defining a succession of trilayers embedded in an environment to be defined self-consistently,
in  manner analogous to that applied the Coherent Phase Approximation\cite{S67si}.
However, we will not focus on this issue in the following and we just assume that the infinite stack identifies a single Moir\'e pattern, with characteristic length
$L=\frac{d}{2\sin\th/2}\simeq d/\th$, where
$d=2.46\,\text{\AA}$ is the lattice constant of monolayer graphene.
This allows to define a reciprocal lattice generated by the two basis vectors
\bea\lb{recvec}
 \mathbf{G}_1=\frac{4\pi}{\sqrt{3}L}(1/2,\sqrt{3}/2)\text{ and }
  \mathbf{G}_2=\frac{4\pi}{\sqrt{3}L}(-1,0).
 \eea
 The corresponding Brillouin zone (BZ) is shown in Fig.~\ref{BZ}, left,
 where we have $\mathbf{K}\equiv 4\pi(0,1)/(3L)$ and
 $\mathbf{M}\equiv \pi(1,\sqrt{3})/\left(\sqrt{3}L\right)$. It is interesting to note that the model has a non-symmorphic symmetry, as a translation by one layer, followed by a twist, leaves the Hamiltonian invariant. The calculations are done using a unit cell which contains three layers, see below. This implies that a new, smaller, Brillouin Zone can be defined, see Fig.~\ref{BZ}, right. We will follow the convention used in the study of twisted bilayers, and use the reciprocal lattice vectors $\mathbf{G}_1,\mathbf{G}_2$ of Eq.~\pref{recvec}. When studying its spectrum, we will use the high-symmetry points in the small triple-layer Brillouin Zone, shown on the right in Fig.~\ref{BZ}. It is worth noting that the same situation occurs in rhombohedral graphite which has a $ABCABC \cdots$ stacking.
 \begin{figure}[ht!]
 \begin{center}
 \begin{tabular}{lr}
\includegraphics[scale=0.25]{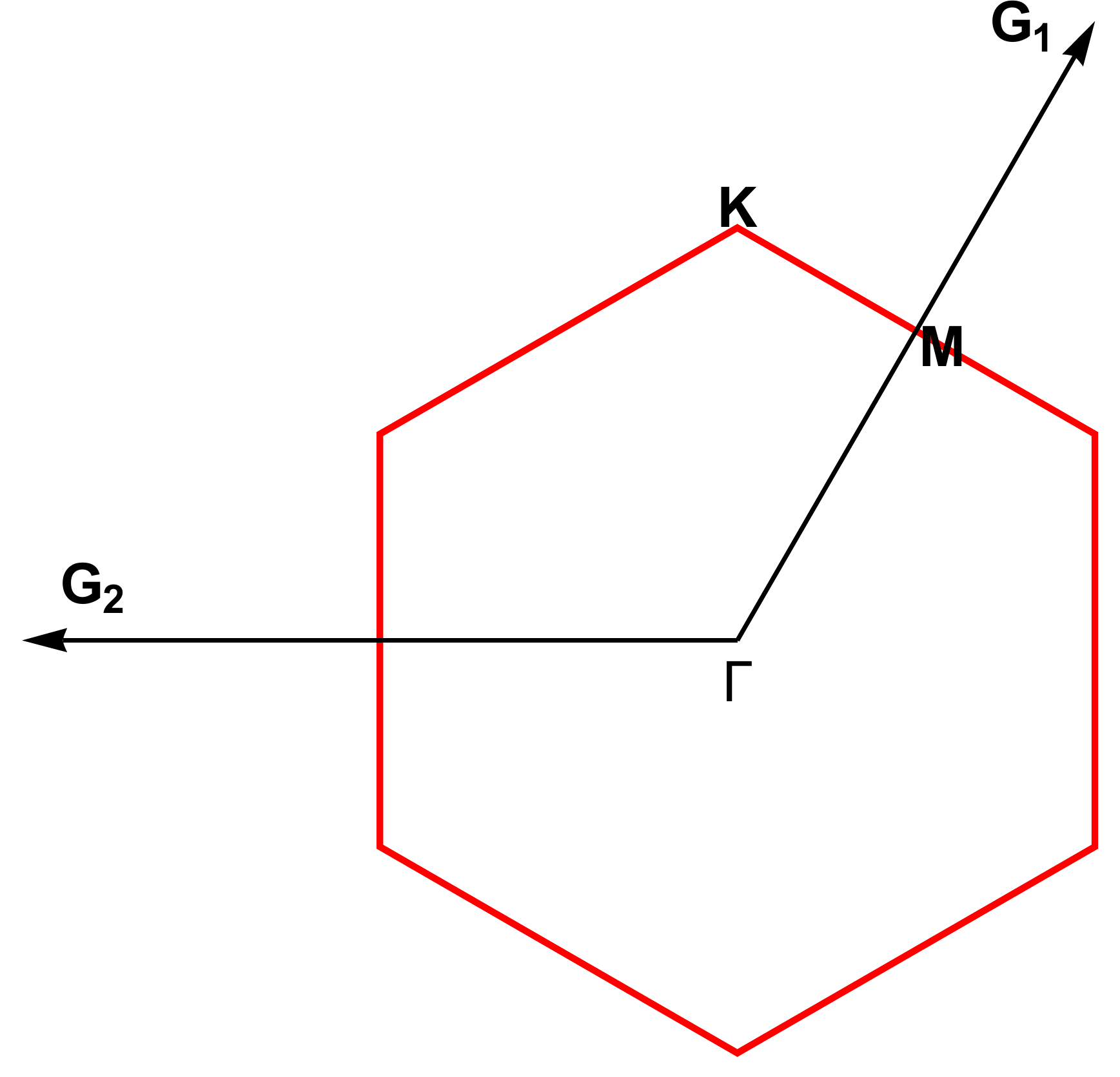} &
\includegraphics[scale=0.225]{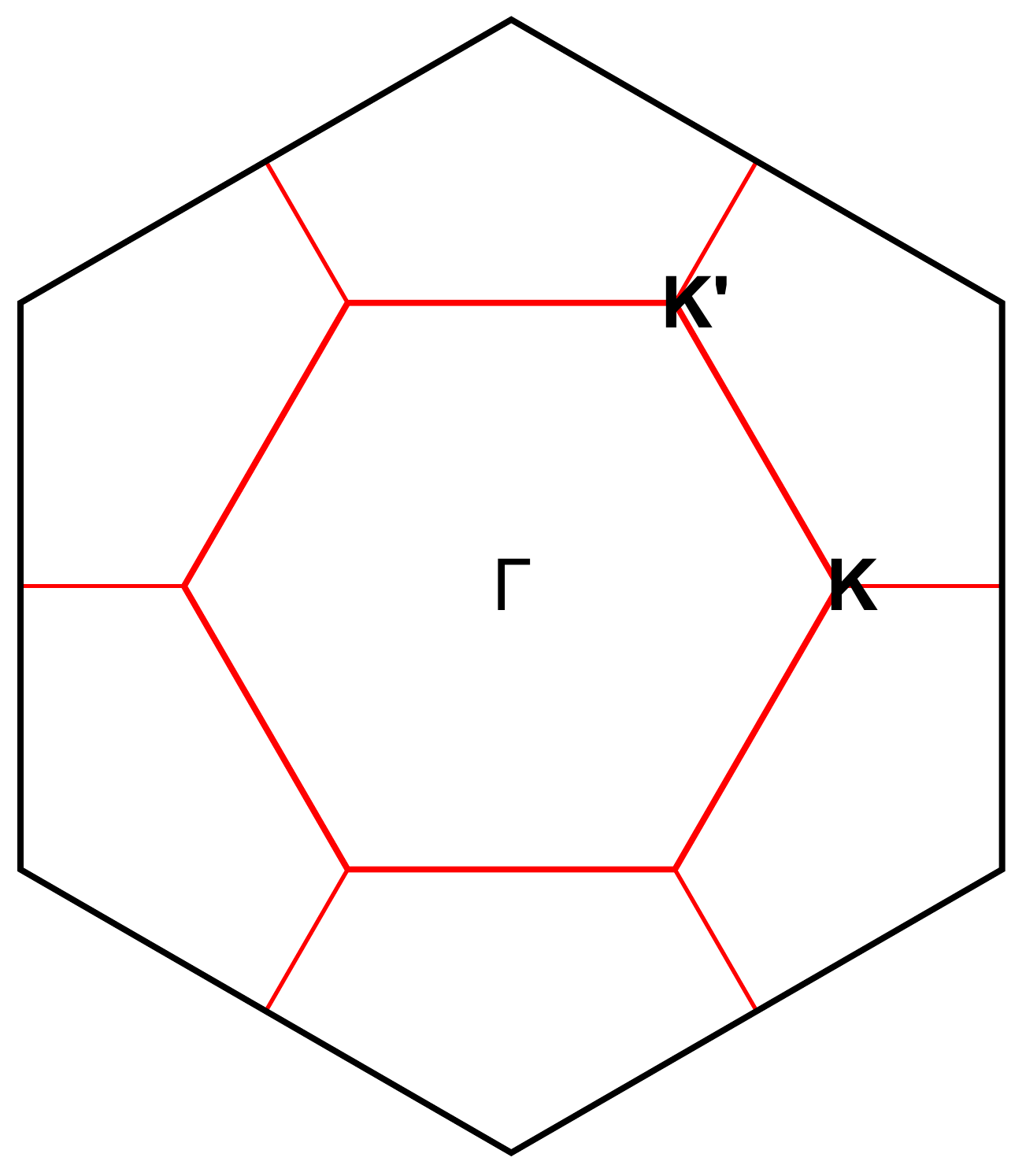}
\end{tabular}
\end{center}
\caption{
Brillouin zone of the Moir\'e lattice. Left: Brillouin Zone defined for each graphene layer. Right: smaller Brillouin Zone defined for three twisted layers.
}\lb{BZ}
\end{figure}

The basic building block for the model presented here is  twisted
trilayer graphene, and we built an infinite system by stacking these on top of each other. The top layer of each block is coupled to the
bottom layer of the next one,
so that neighboring layers of consecutive blocks form a twisted bilayer with the same twist angle.
Each block can be modeled within the continuum framework first introduced in
\cite{santos_prl07,Bistritzer_pnas11,santos_prb12} for the twisted bilayer and
recently extended to the twisted trilayer in Ref.~\cite{mora_cm19}.
Within this continuum model, the two valleys of each graphene layer at $K$ and $K'$ can be treated separately.
This approximation is still valid for the case of bulk three-dimensional systems considered here,
where inter-valley effects occur only at very high energies
\cite{mcclure_physrev57,Slonczewski_physrev58,castroneto_revmodphys09},
and hence they do no not affect the band structure close to the Dirac points.
Consequently, in the following we will focus only on the $K$ valley of each layer.
Using the notation of \cite{santos_prl07,santos_prb12}, we write the Hamiltonian of the $n$-th block as
\bea\lb{Hnn}
H_{nn}&=&\int\,d^2k\left\{
\sum_{l=1}^3v_F\Psi^\dagger_{n,l}(\bk)\vec{\sigma}\cdot\left[\bk-\left(l-2\right)\mathbf{K}\right]\Psi_{n,l}(\bk)+\right.\\
&+&\left.
\sum_{\mathbf{G}}\left[
 \Psi^\dagger_{n,1}(\bk) \hat{T}_\perp(\mathbf{G}) \Psi_{n,2}(\bk-\mathbf{G})+
  \Psi^\dagger_{n,2}(\bk) \hat{T}_\perp(\mathbf{G}) \Psi_{n,3}(\bk-\mathbf{G})+
 \text{h.c.}
\right]
\right\},\nn
\eea
where $n,l$ are the block and layer indices, respectively,
$v_F=d\g_0\sqrt{3}/2$ is the Fermi velocity of monolayer graphene,
and $\g_0=2.47\,\text{eV}$ \cite{moon_prb13,koshino_PRX18}
is the nearest-neighbor hopping amplitude. The wavefunction
$\Psi^T_{n,l}\equiv(\psi^A_{n,l},\psi^B_{n,l})$,
with $A,B$ the sub-lattice indices, 
$\vec{\s}\equiv(\s_x,\s_y)$ are the Pauli matrices,
$\mathbf{G}=n\mathbf{G}_1+m\mathbf{G}_2$, $n,m$ integers.
Finally
$\hat{T}_\perp(\mathbf{G})$ is the $2\times2$ matrix which parametrizes
the hopping between consecutive layers.
To a first approximation we consider only the contribution of the first three components
of $\hat{T}(\mathbf{G})$:
$\hat{T}_\perp(0),\hat{T}_\perp(-\mathbf{G}_1),\hat{T}_\perp(-\mathbf{G}_1-\mathbf{G}_2)$,
with the corresponding matrix elements given by (see \cite{santos_prl07,koshino_PRX18})
\bea
\hat{T}_\perp(0)=\begin{pmatrix}g_1&&g_2\\g_2&&g_1\end{pmatrix}\quad,
\quad
\hat{T}_\perp(-\mathbf{G}_1)=
\begin{pmatrix}g_1e^{2i\pi/3}&&g_2\\g_2e^{-2i\pi/3}&&g_1e^{2i\pi/3}\end{pmatrix}=
\left[\hat{T}_\perp(-\mathbf{G}_1-\mathbf{G}_2)\right]^*.
\eea
We shall employ the values recently obtained in \cite{koshino_PRX18}:
$g_1=0.0797$eV, $g_2=0.0975$eV.
The difference between $g_1$ and $g_2$,
not considered in the former models of twisted bilayer graphene,
introduces energy gaps between the lowest bands and the excited bands,
in qualitative agreement with the experimental results \cite{cao_prl16,cao_nat18,cao_nat18_2}.

Notice that, in the absence of the interlayer interaction, the Hamiltonian
$H_{nn}$ gives rise to Dirac cones at the point $\pm\mathbf{K},\Gamma$ of the BZ.
It is also worth noting that in Eq.~\pref{Hnn} we are neglecting the small rotation
of the pseudo-spin $\vec{\sigma}$, which arises due to the
twist between consecutive layers.
This approximation has been shown to be valid for the case of trilayer graphene in Ref.~\cite{mora_cm19}; 
we assume that it still holds in the case of infinite layers, as long as we can identify a Moir\'e Broillouin zone, i.e.,  for small twist angles.

The Hamiltonian coupling consecutive blocks can been written as
\bea\lb{Hnn+1}
H_{n,n+1}=\int\,d^2k\sum_\mathbf{G}
 \Psi^\dagger_{n,3}(\bk) \hat{T}_\perp(\mathbf{G})
 \Psi_{n+1,1}(\bk-\mathbf{G}-2\mathbf{G}_1-\mathbf{G}_2)+\text{\text{h.c.}}\quad,
\eea
and describes the hopping between the third layer of the $n$-th block
and the first layer  of the $(n+1)$-th block. 

We note that the Dirac points of the $n$-th block are reciprocally connected to the ones of the $n+1$-th block by the reciprocal lattice vector: $2\mathbf{G}_1+\mathbf{G}_2$, which is why the same vector appears in the argument of $\Psi_{n+1,1}$ in the rhs of the Eq. \pref{Hnn+1}. Such a situation occurs only if the building blocks of the stack are arranged in multiples of three layers, which therefore justifies our choice of partitioning the stack in such blocks.

The full Hamiltonian is then
$$
H=\sum_{n=-\infty}^{\infty}\left(H_{n,n}+H_{n,n+1}\right)\,,
$$
and can be easily diagonalized by exploiting its translational invariance along the $z$ axis
perpendicular to the layers by Fourier transformation.
In term of the momentum $k_z$ we find
\bea
H&=&\int_{-{\pi}/{3c}}^{{\pi}/{3c}}\,dk_z H_{k_z},\lb{Hfull}\\
H_{k_z}&=&\int\,d^2k\left\{
\sum_{l=1}^3v_F\Psi^\dagger_{k_z,l}(\bk)\vec{\sigma}\cdot\left[\bk-\left(l-2\right)\mathbf{K}\right]\Psi_{k_z,l}(\bk)+
\right.\lb{Hkz}\\
&&+\left.
\sum_{\mathbf{G}}\left[
 \Psi^\dagger_{k_z,1}(\bk) \hat{T}_\perp(\mathbf{G}) \Psi_{k_z,2}(\bk-\mathbf{G})+
  \Psi^\dagger_{k_z,2}(\bk) \hat{T}_\perp(\mathbf{G}) \Psi_{k_z,3}(\bk-\mathbf{G})+\right.\right.\nn\\
  &&\qquad +\left.\left.
  e^{i3k_z c} \Psi^\dagger_{k_z,3}(\bk) \hat{T}_\perp(\mathbf{G})
 \Psi_{k_z,1}(\bk-\mathbf{G}-2\mathbf{G}_1-\mathbf{G}_2)+
 \text{\text{h.c.}}
\right]
\right\},\nn
\eea
where $c=3.35${\AA} is the interlayer spacing and $\Psi_{k_z,l}(\bk)\equiv\sqrt{\frac{3c}{2\pi}}\sum_n\Psi_{n,l}(\bk)e^{-ik_z3cn}$.

The interlayer interaction of the Hamiltonian \pref{Hkz} then
couples states of layer 1 (2) with momentum $\bk$ to states of layer 2 (3) with momentum
$\bk$, $\bk+\mathbf{G}_1$ and $\bk+\mathbf{G}_1+\mathbf{G}_2$,
and states of layer 3 with momentum $\bk$ to states of layer 1 with momentum
 $\bk-2\mathbf{G}_1-\mathbf{G}_2$, $\bk-\mathbf{G}_1-\mathbf{G}_2$ and
 $\bk-\mathbf{G}_1$.

The Hamiltonian \pref{Hkz} can be diagonalized in the reciprocal space
by looking for the eigenfunctions as a superposition of Bloch waves,
\bea
\Phi_{k_z}(\bk,\mathbf{r})=\sum_{nm}
\begin{bmatrix}
\phi^{A1}_{nm,k_z}(\bk,\mathbf{r})\\
\phi^{B1}_{nm,k_z}(\bk,\mathbf{r})\\
\phi^{A2}_{nm,k_z}(\bk,\mathbf{r})\\
\phi^{B2}_{nm,k_z}(\bk,\mathbf{r})\\
\phi^{A3}_{nm,k_z}(\bk,\mathbf{r})\\
\phi^{B3}_{nm,k_z}(\bk,\mathbf{r})
\end{bmatrix}\times
e^{i\left(\bk+n\mathbf{G}_1+m\mathbf{G}_2\right)\cdot\mathbf{r}},
\eea
where $\bk$ belongs to the Moir\'e BZ shown in Fig.~\ref{BZ}.
We perform the numerical diagonalization by truncating the Bloch waves expansion
at momenta large enough to achieve
the convergence of the lowest energy bands.
The results shown in the Fig.~2 of the main text and in the following
have been obtained by considering a set
of 169 Bloch states.

In Fig.~\ref{inf_layers_bands} we show the bands and
the density of the states (DOS) for a few additional twist angles, $\th=0.82^\circ,0.55^\circ$.
The red line superimposed to the DOS represents the case of three uncoupled layers.
 \begin{figure}[ht!]
\includegraphics[width=5in]{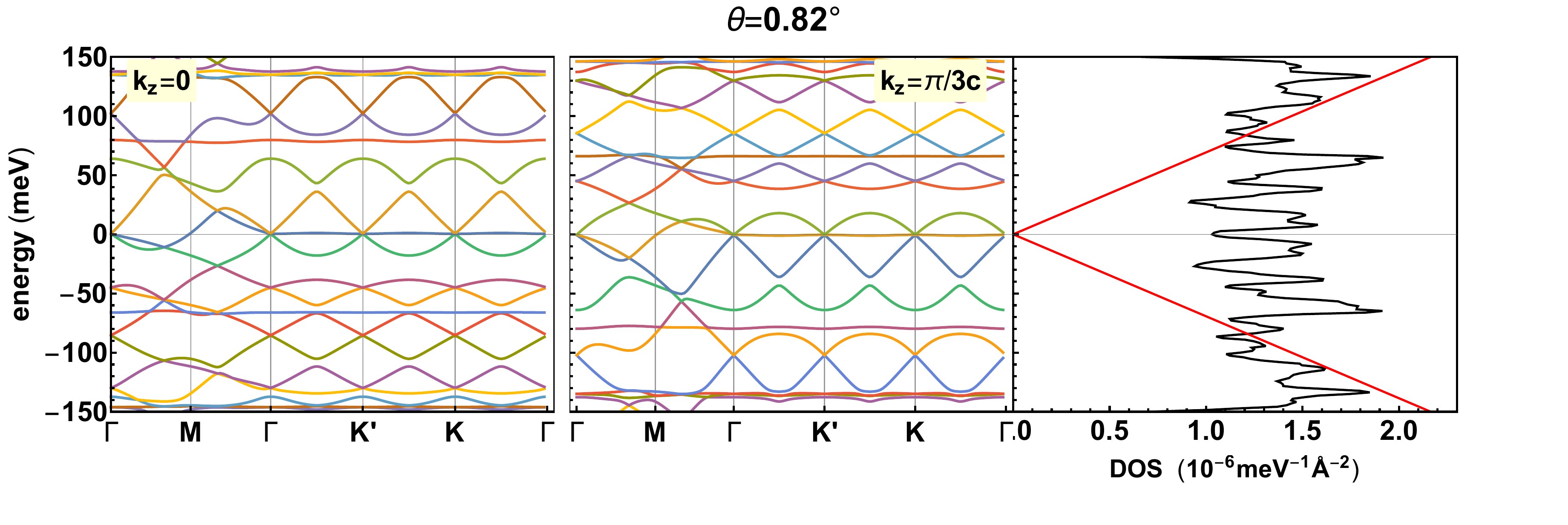}\\
\includegraphics[width=5in]{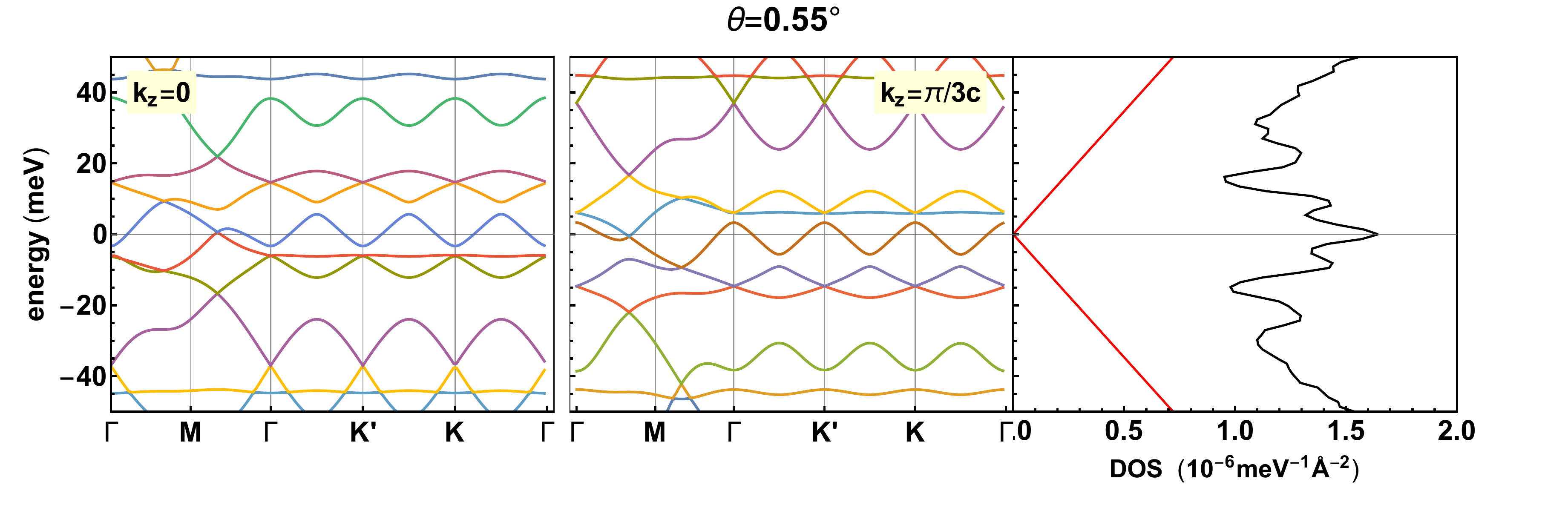} 
\caption{Bands and DOS of an infinite stack of twisted graphene layers for $\th=0.82^\circ,0.55^\circ$.
The red line superimposed on the DOS shows the graphene dispersion for three uncoupled layers.}
\lb{inf_layers_bands} 
\end{figure}

In Fig.~\ref{charge_density} we show the charge-density distribution for the
three lowest energy bands obtained for the angle $\th=1.08^\circ$, at $k_z=0$ and
at the points $\Gamma$ (left), $K$ (central), $K'$ (right) of the small Moir\'e BZ.
The middle  row, which corresponds to the flat band, shows that the charge density is zero in the $AA$ region
(see the schematic representation of the layer alignment in the unit cell in the bottom panel).
The central band shows similar charge distributions for the three high symmetry points, while the neighboring bands show inequivalent distributions for each of the three points. This situation is different from that found in a twisted bilayer, where the $K$ and $K'$ points are equivalent, and different from the $\Gamma$ point\cite{RM18si}. We do not explore here how these symmetries influence the Wannier functions which describe these bands\cite{koshino_PRX18}.

 \begin{figure}[ht!]
 \begin{tabular}{ccc}
\includegraphics[width=2in]{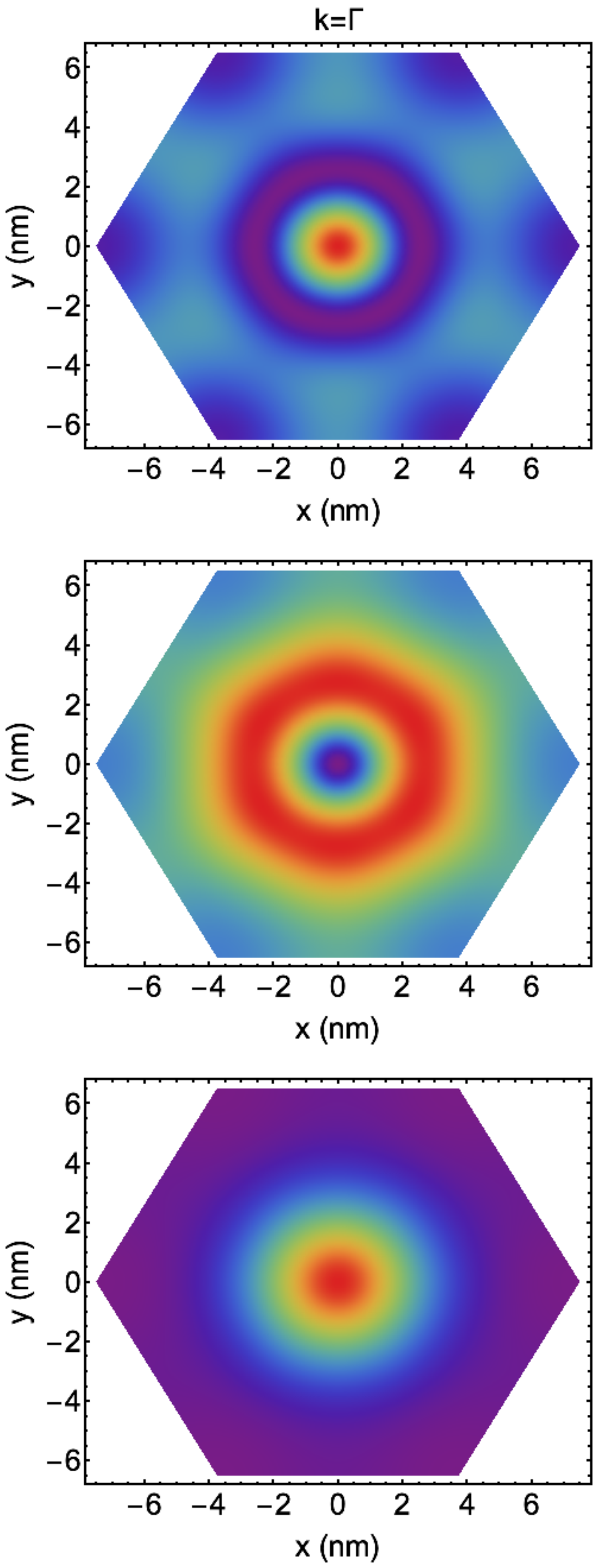}&
\includegraphics[width=2in]{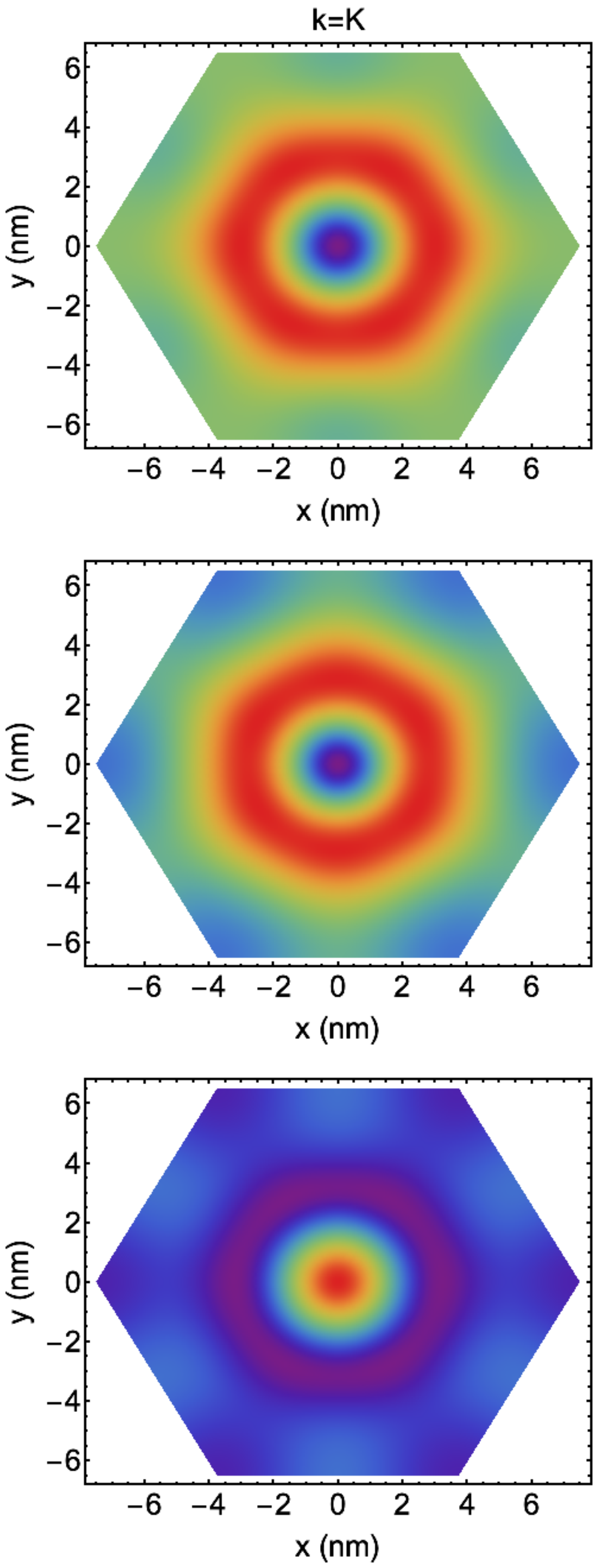} &
\includegraphics[width=2in]{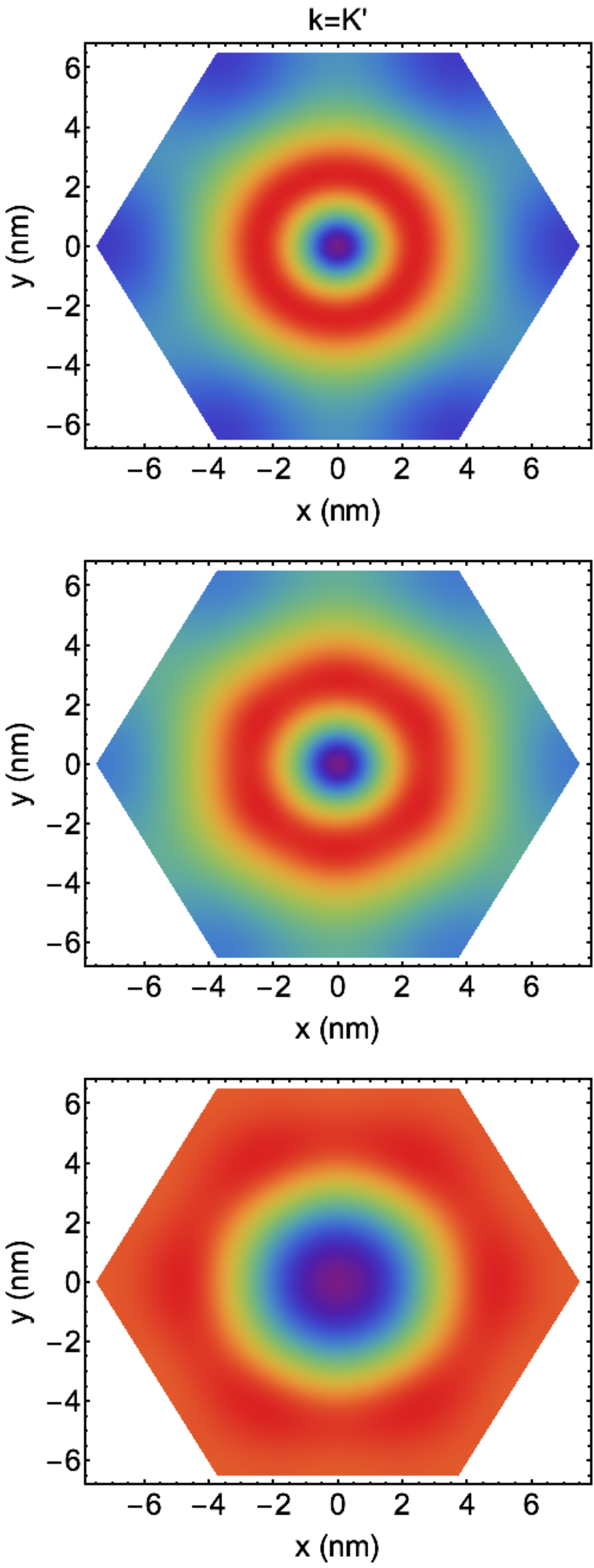}
\end{tabular}
\includegraphics[width=2.in]{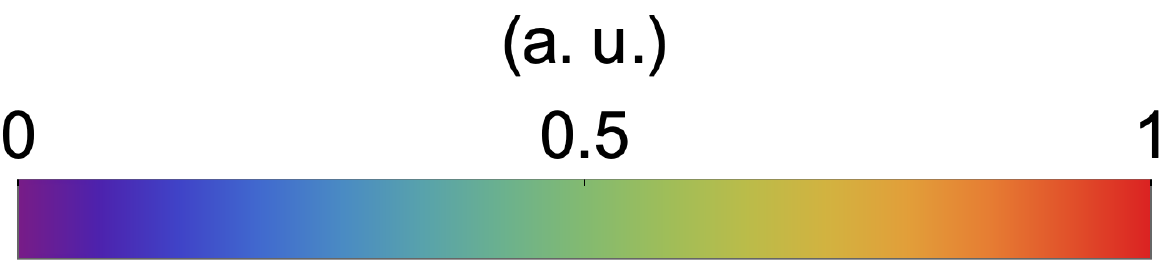}\\
\includegraphics[width=1.5in]{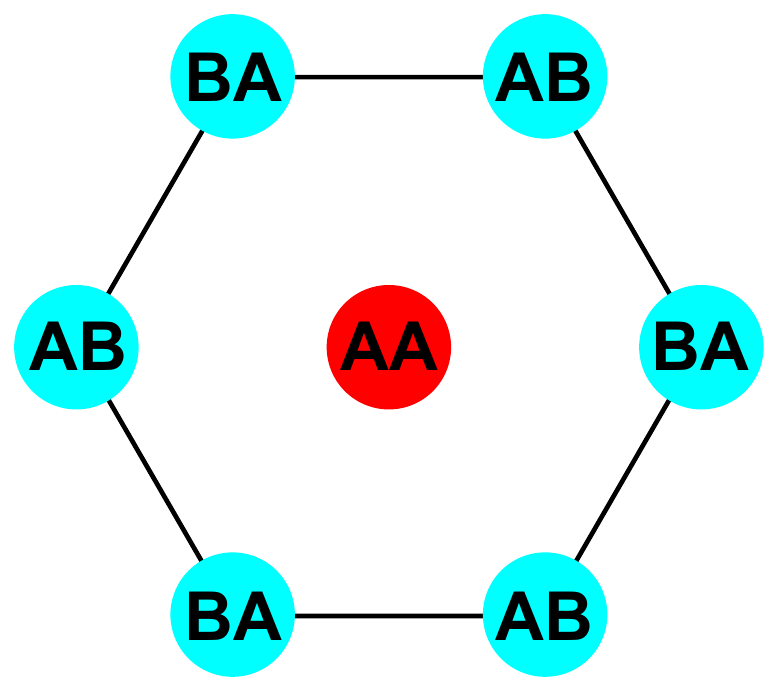} 
\caption{
Charge density distribution of the
three lowest-energy bands obtained for the twist angle $\th=1.08^\circ$,
at $k_z=0$ and 
at the points $\Gamma$ (left panels), $K$ (central panels), $K'$ (right panels) of the small BZ,
in the Moir\'e unit cell.
The bottom panel shows a schematic representation of the layer alignment across  the unit cell.
}
\lb{charge_density}
\end{figure}

\section{Twisted Stacking Fault in Bulk Graphite}

In contrast to the previous section,
where we considered an infinite stack of graphene
layers with a constant twist between a layer and the consecutive one,
here we study the case of one single twist between two semi-infinite
stacks placed one above the other, as shown in the Fig. 1(b) of the main text.
Every stack is assumed to be of Bernal type.
This situation is indeed particularly meaningful for describing defects in graphite.

We neglect hopping between
next nearest neighbor layers in different layers across the stacking defect,
so that the coupling between the two stacks is restricted to the layers at the interface.
Our approach consists in studying the twisted bilayer represented by
the bottom layer of the upper ($U$) stack and the top layer of the lower ($L$) stack,
once the interior (core) layers have been integrated out.
We describe this bilayer within the
path integral formulation, by means of the following effective action
\bea
S_{\text{eff}}&=&\int\,d\o\, \mathcal{L}_{\text{eff}}(\o) \\
 \mathcal{L}_{\text{eff}}(\o)&=&\int\,d^2k \sum_{i=L,U} \Psi^\dagger_i(\bk,\o)G^{-1}_i(\bk,\o) \Psi_i(\bk,\o)-
H_{LU},\lb{Seffom}
\eea
where $\Psi^T_i=(\psi^A_i,\psi^B_i)$ is the spinor for the layer placed at the surface of the stack $i$,
$G_i$ is the corresponding effective Green's function,
which encodes the information concerning the spectrum of graphite,
and $H_{LU}$ is the interlayer Hamiltonian of a twisted bilayer\cite{santos_prl07,santos_prb12},
\bea
H_{LU}=\int\,d^2k\sum_{\mathbf{G}}
 \Psi^\dagger_L(\bk,\o) \hat{T}_\perp(\mathbf{G}) \Psi_U(\bk-\mathbf{G},\o)+\text{h.c.}
\eea
We investigate the spectrum of the system by computing
the DOS per unit of momentum $\rho(\bk,\o)$,
which is defined by the eigenvalues $\xi_{\bk,\a}(\o)$ of $ \mathcal{L}_{\text{eff}}(\o)$ according to
\bea\lb{DOS}
\rho(\bk,\o)=-\frac{1}{\pi}\sum_\a\mathrm{Im}\left\{\xi_{\bk,\a}(\o+i\d)^{-1}\right\},
\eea
where $\a$ is the band index and $\d>0$ is a finite spectral broadening.

In order to describe the procedure for computing the Green's function $G_i$,
in the following we focus on the upper stack $i=U$, without loss of generality.
We consider all its sub-stacks $\left(BA\right)_0,\left(BA\right)_1\dots$,
where $0$ denotes the bottom block.
The action of the stack is
\bea\lb{Sstack}
S_{\text{stack}}=\int\,d\o\,d^2k\sum_{n\ge0}\left\{
\Xi^\dagger_n(\bk,\o)\mathcal{G}^{-1}_0(\bk,\o)\Xi_n(\bk,\o)-
\left[
\Xi^\dagger_n(\bk,\o)\hat{T}_0(\bk)\Xi_{n+1}(\bk,\o)+\text{h.c.}
\right]
\right\},
\eea
where
$\Xi^T_n=(\psi_{A1,n},\psi_{B1,n},\psi_{A2,n},\psi_{B2,n})$
is the spinor for the bilayer in the $n$-th sub-stack,
$\mathcal{G}^{-1}_0(\bk,\o)=\o\mathcal{I}-\hat{H}(\bk)$
is the inverse of the non-interacting Green's function.
The operators $\hat{H}(\bk)$, $\hat{T}_0(\bk)$ are the $4\times4$
matrices of the couplings in each sub-stack and
between consecutive sub-stacks, respectively
\cite{Slonczewski_physrev58,castroneto_revmodphys09,
mccann_repprogphys13}:
\bea\lb{4X4Hmat}
\hat{H}(\bk)=
\begin{pmatrix}
0&v_Fk&-v_4k&v_3k^*\\
v_Fk^*&0&\g_1&-v_4k\\
-v_4k^*&\g_1&0&v_Fk\\
v_3k&-v_4k^*&v_Fk^*&0
\end{pmatrix}
\quad,\quad
\hat{T}_0(\bk)=\begin{pmatrix}
\g_2&0&0&0\\
0&-\g_2&0&0\\
-v_4k^*&\g_1&-\g_2&0\\
v_3k&-v_4k^*&0&\g_2
\end{pmatrix}.
\eea
Here $k\equiv (k_x-K'_x)-i(k_y-K'_y)$, the Dirac point lying at $\mathbf{K'}$,
$\g_1$ is the coupling between orbitals in atoms that are
nearest neighbors in consecutive layers,
$v_{3,4}\equiv v_F\g_{3,4}/\g_0$, where $\g_3,\g_4$
are the hopping amplitudes between orbitals at
next nearest neighbors in consecutive layers,
and $\g_2$ is the coupling between orbitals at next nearest neighbors layers.
Notice that we are neglecting the parameter
$\Delta$ and we are assuming $\g_5=\g_2$.
The values of the parameters
$\g$ in graphite have been intensively studied
(see\cite{mcclure_physrev57,nozieres_physrev58,
dresselhaus_IBM64,soule_physrev64,
dillon_JPCS77,brandt_book88}).
In the following we refer to those reported in \cite{dresselhaus_physrev140}:
$\g_0=3.16$eV, $\g_1=0.39$eV, $\g_2=-0.02$eV, $\g_3=0.315$eV, $\g_4=0.044$eV.

Integrating the sub-stacks $1,\dots N$ out of the action \pref{Sstack},
the effective action of the remaining sub-stacks can be written as:
\bea
S_{\text{stack}}^{(N)}&=&\int\,d\o\,d^2k\biggl\{
\Xi^\dagger_0(\bk,\o)\tilde{\mathcal{G}}^{-1}_N(\bk,\o)\Xi_0(\bk,\o)+
\Xi^\dagger_{N+1}(\bk,\o)\mathcal{G}^{-1}_N(\bk,\o)\Xi_{N+1}(\bk,\o)\\
&&\qquad-\left.
\left[\Xi^\dagger_0(\bk,\o)\hat{T}_N(\bk,\o)\Xi_{N+1}(\bk,\o)+\text{h.c.}\right]\right.\nn\\
&&\qquad+
\sum_{n> N+1}\Xi^\dagger_n(\bk,\o)\mathcal{G}^{-1}_0(\bk,\o)\Xi_n(\bk,\o)\nn\\
&&\qquad-
\sum_{n\ge N+1}\left[ \Xi^\dagger_n(\bk,\o)\hat{T}_0(\bk)\Xi_{n+1}(\bk,\o)+\text{h.c.}\right]
\biggr\}\nn,
\eea
where $\tilde{\mathcal{G}}_N$
is the effective Green's function of the bottom sub-stack,
$\mathcal{G}_N$ that of the $(N+1)$-th
sub-stack and $\hat{T}_N$ is the effective
coupling between them. It can be easily shown that
$\tilde{\mathcal{G}}_N,\mathcal{G}_N$ and $\hat{T}_N$
are obtained from the $(N-1)$-th step according to the recursive relations:
\bea
\tilde{\mathcal{G}}^{-1}_N&=&
\tilde{\mathcal{G}}^{-1}_{N-1}-
\hat{T}_{N-1}\mathcal{G}_{N-1}\hat{T}^\dagger_{N-1},\nn\\
\mathcal{G}^{-1}_N&=&\mathcal{G}^{-1}_0-
\hat{T}_0^\dagger \mathcal{G}_{N-1}\hat{T}_0,\lb{iter_GN}\\
\hat{T}_N&=&\hat{T}_{N-1}\mathcal{G}_{N-1}\hat{T}_0,\nn
\eea
with $N>0$ and $\tilde{\mathcal{G}}_0=\mathcal{G}_0$.
Because the inter-layer hopping amplitudes $\g_1\dots\g_4$
are smaller than the intra-layer one $\g_0$, then $\hat{T}_N\to0$
as $N\to\infty$.
Thus $\tilde{\mathcal{G}}_N$ converges to a fixed point $\tilde{\mathcal{G}}_\infty$,
which describes the dynamics of the bottom
sub-stack accounting for all the interactions with the upper ones.
This method can of course also be used to study a finite stack with $2N$ layers.

The $2\times2$ inverse Green's function $G^{-1}_U$ introduced in Eq.~\pref{Seffom}
is the projection of $\tilde{\mathcal{G}}^{-1}_\infty$ on the first $2\times2$ block,
meaning that, if
$$
\tilde{\mathcal{G}}^{-1}_\infty=
\begin{pmatrix}
\mathcal{G}_{11}^{-1}&&\mathcal{G}_{12}^{-1}\\
\mathcal{G}_{12}^{-1,\dagger}&&\mathcal{G}_{22}^{-1}
\end{pmatrix}
\quad,
\quad\text{then}\quad
G^{-1}_U=\mathcal{G}_{11}^{-1}-
\mathcal{G}_{12}^{-1}\mathcal{G}_{22}\mathcal{G}_{12}^{-1,\dagger}.
$$
The procedure for obtaining $G^{-1}_L$ is equivalent to that described above.
The only changes are the swap of the indices $1\leftrightarrow2$ and $3\leftrightarrow4$ and $k\leftrightarrow k^*$
in the matrices of the Eq.~\pref{4X4Hmat}, as the stack is now arranged as $(AB)_0,(AB)_1,\dots$.

The spectra in Fig.~3 of the main text show the DOS, $\rho(\bk,\o)$, given by the
Eq. \pref{DOS} and obtained by truncating the iteration \pref{iter_GN} at
$N=50$, where convergence can be considered reached.
We use the spectral broadening $\d=10^{-3}\g_0$.

Finally we note that in the simplified case where
$\g_2=\g_3=\g_4=0$ while $\g_1$ is finite,
 $G^{-1}_i$ takes the following analytic form:
\bea\lb{GLU_analytic}
G^{-1}_L(\bk,\o)=\begin{pmatrix}
\o-\Sigma(\bk,\o)&&-v_Fk\\-v_Fk^*&&\o
\end{pmatrix}\quad,\quad
G^{-1}_U(\bk,\o)=\begin{pmatrix}
\o&&-v_Fk\\-v_Fk^*&&\o-\Sigma(\bk,\o)
\end{pmatrix},
\eea
where
\bea
\Sigma(\bk,\o)=\frac{\o\g_1^2}{\o^2-\left(v_Fk\right)^2}\left[1+
\sum_{n\ge1}
\frac{\prod_{l=1}^n\left(\g_1 a_l\right)^2}{1-\frac{\o a_n\g_1^2}{\o^2-\left(v_Fk\right)^2}}
\right],
\eea
and the coefficients $a_n$ satisfy the recursive equation:
\bea
a_n=\frac{\o}{\o^2-\left(v_Fk\right)^2-\o\g_1^2a_{n-1}}\quad,\quad a_0=0.
\eea
We show  the DOS resulting from this analytic Green's functions \pref{GLU_analytic}
for the twist angle $\th=1.08^\circ$ in Fig.~\ref{DOS_analytic}.
The isolated flat bands appearing at zero energy are due to the absence of the $\g_2,\g_3,\g_4$ couplings.
 \begin{figure}[ht!]
\includegraphics[scale=0.35]{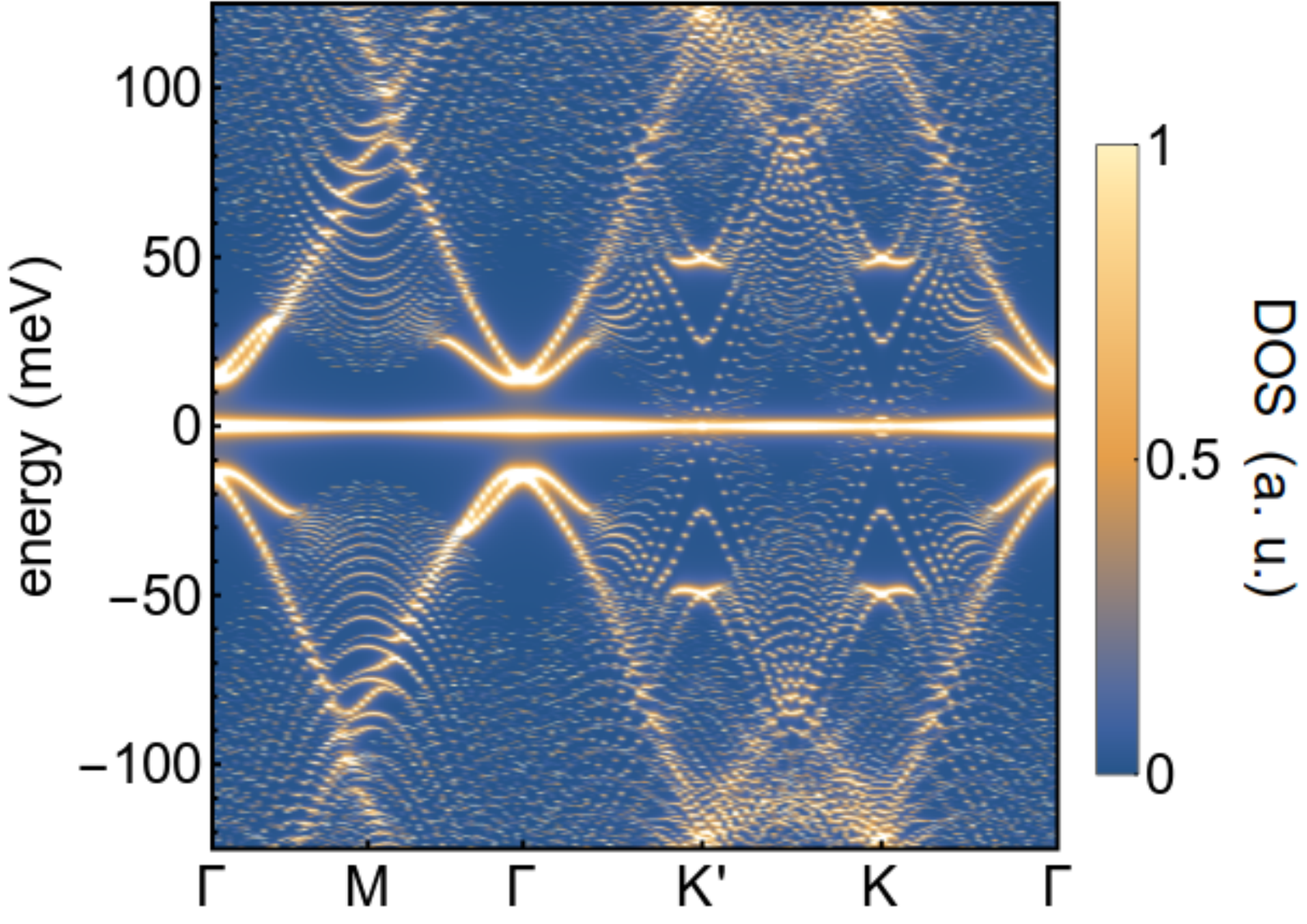}
\caption{
DOS obtained within the analytic result Eq.~\pref{GLU_analytic}
for a twist angle $\th=1.08^\circ$.
}\lb{DOS_analytic}
\end{figure}

\section{Twisted Layer on a Graphite Surface}
The formalism developed in the previous section can be easily adapted to describe
the situation in which a single layer of graphene is rotated with respect to bulk graphite. We only need to compute  $G^{-1}_L$ as above, and can use the graphene result for the other:
$$
G^{-1}_U(\bk,\o)=G^{-1}_0(\bk,\o)=\o-v_F\vec{\s}\cdot\bk\,.
$$
This is then substituted into  Eq.~\pref{Seffom}. 
The resulting spectra are shown in the Fig.~4 of the main text. In  Fig.~\ref{DOS_surface} we show the local DOS at the surface between graphite and the monolayer,
integrated over a narrow energy window of $40\,\text{meV}$ centered at the Fermi level,
at the points $\G$ (left) and $K$ (right) of the Brillouin zone.
 \begin{figure}[ht!]
\begin{tabular}{cc}
\includegraphics[width=3.in]{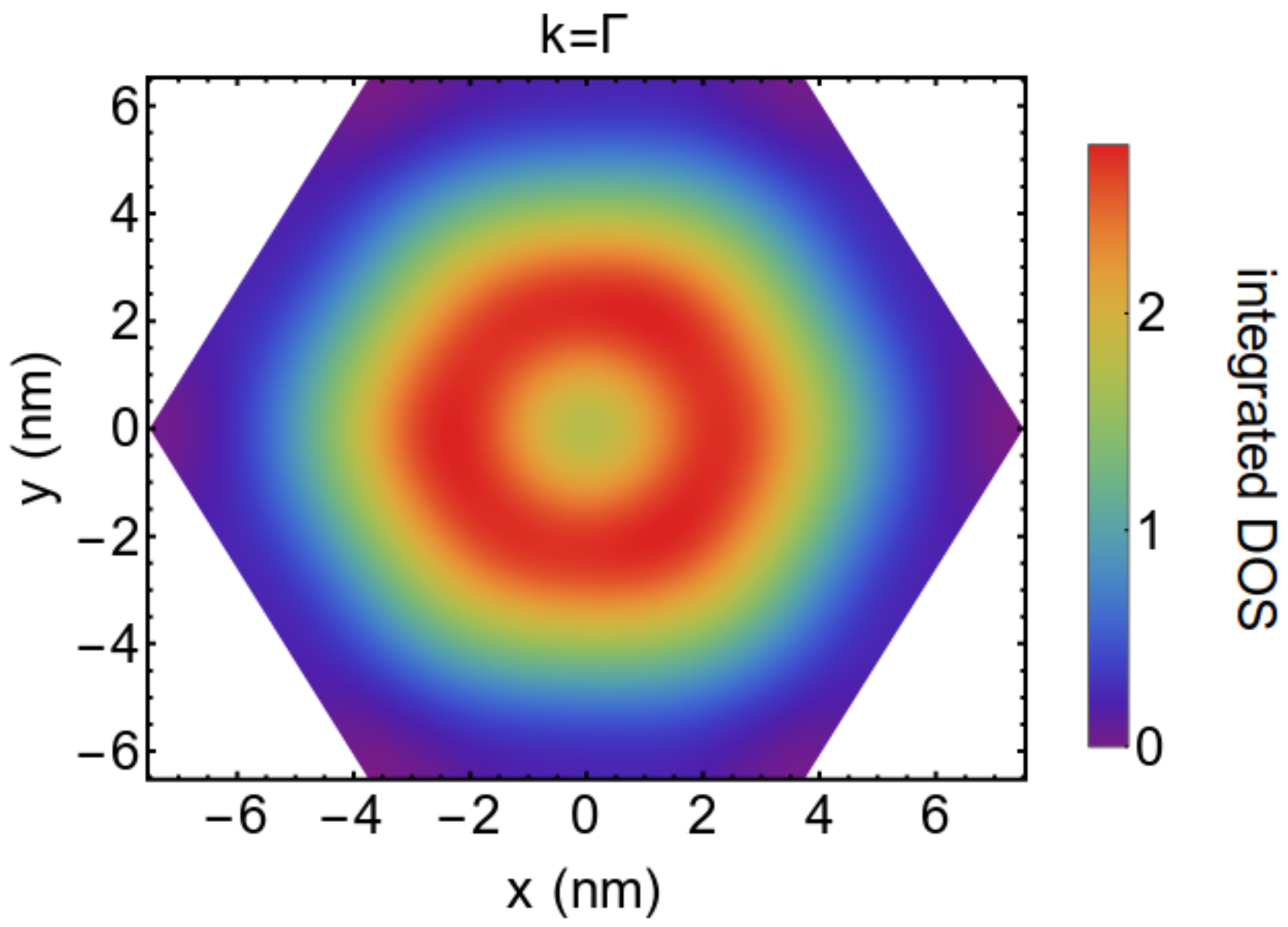} &
\includegraphics[width=3.in]{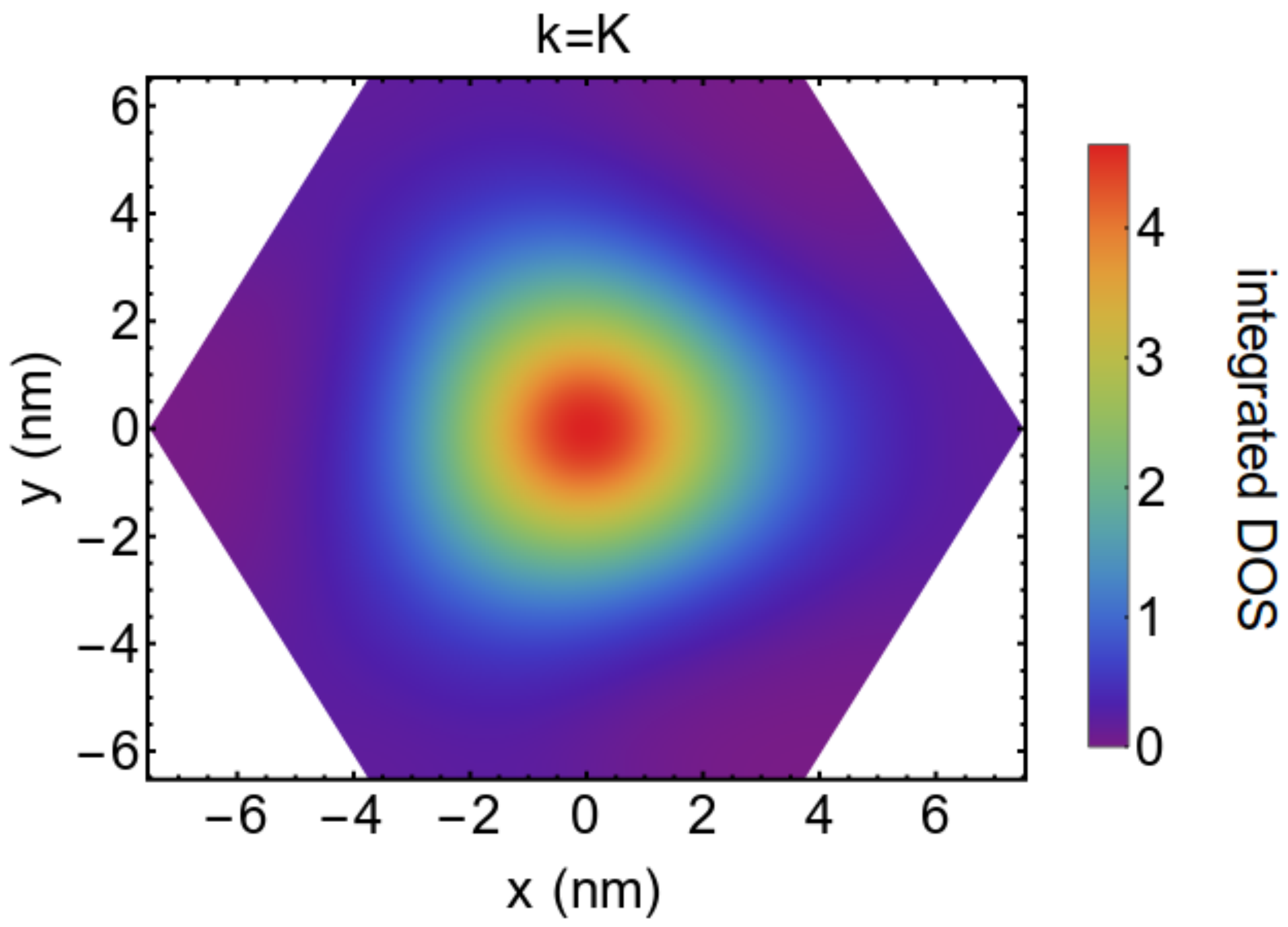} 
\end{tabular}
\caption{
Local DOS at the surface between graphite and the monolayer,
integrated over a narrow energy window of $40\,\text{meV}$ centered at the Fermi level,
at the points $\G$ (left panel) and $K$ (right panel) of the BZ, in the unit cell.}
\lb{DOS_surface}
\end{figure}
These distributions resemble the case for twisted bilayer graphene\cite{RM18si}. They suggest that the symmetries present in twisted bilayer graphene are at least approximate symmetries in this system.

We can also apply a voltage at the interface between graphite and the twisted layer, by putting a gate above the top layer. 
We include the effect of the gate by means of introducing two different chemical potentials,
$\mu_L\ne \mu_U$, in $G^{-1}_L$ and $G^{-1}_U$, respectively.
Figure~\ref{gate} shows the spectra obtained for $\{\mu_L,\mu_U\}=\{5,10\},\{5,20\},\{5,30\}$meV.
 \begin{figure}[ht!]
\begin{tabular}{cccc}
\includegraphics[width=2in]{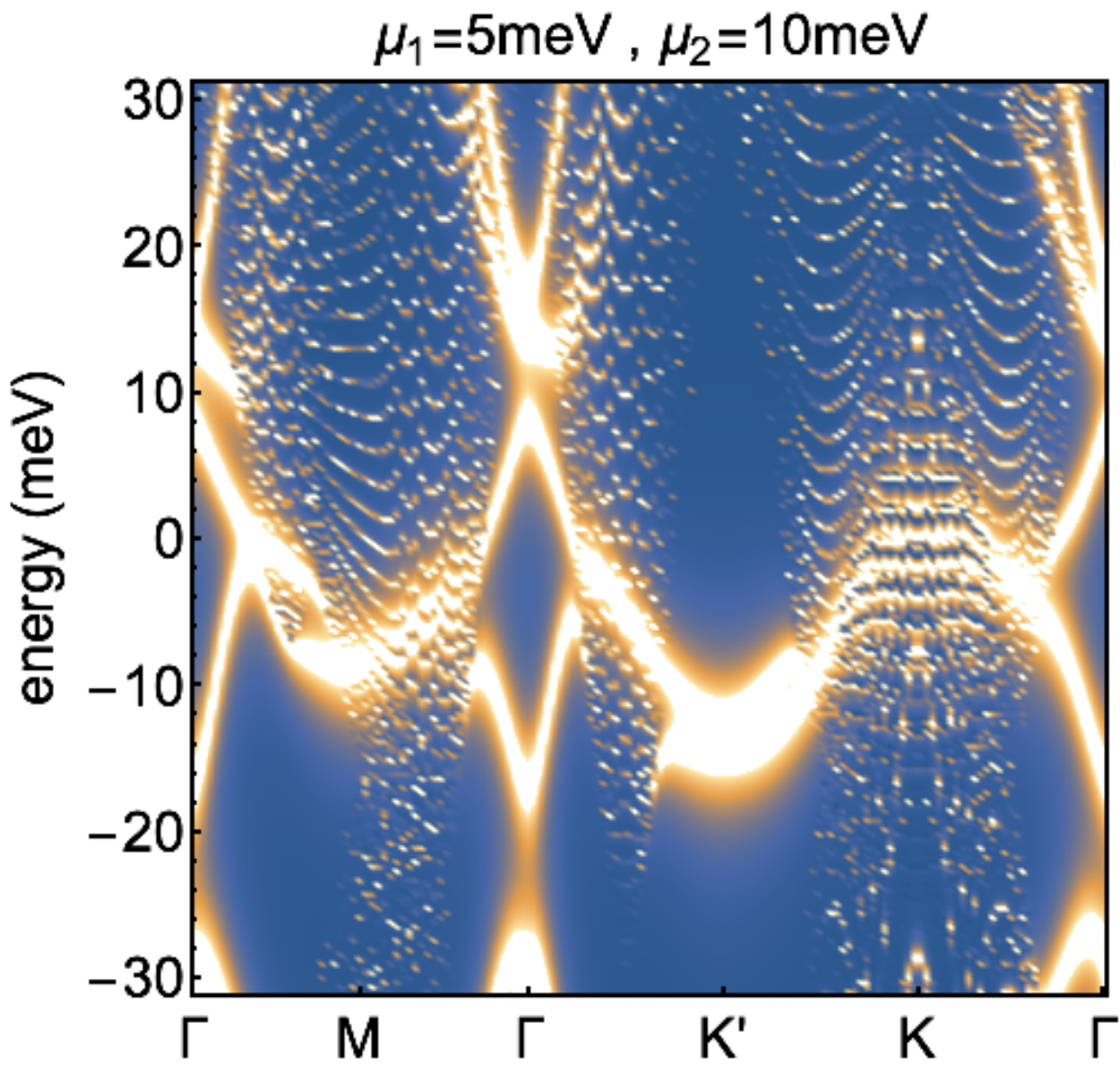} &
\includegraphics[width=2in]{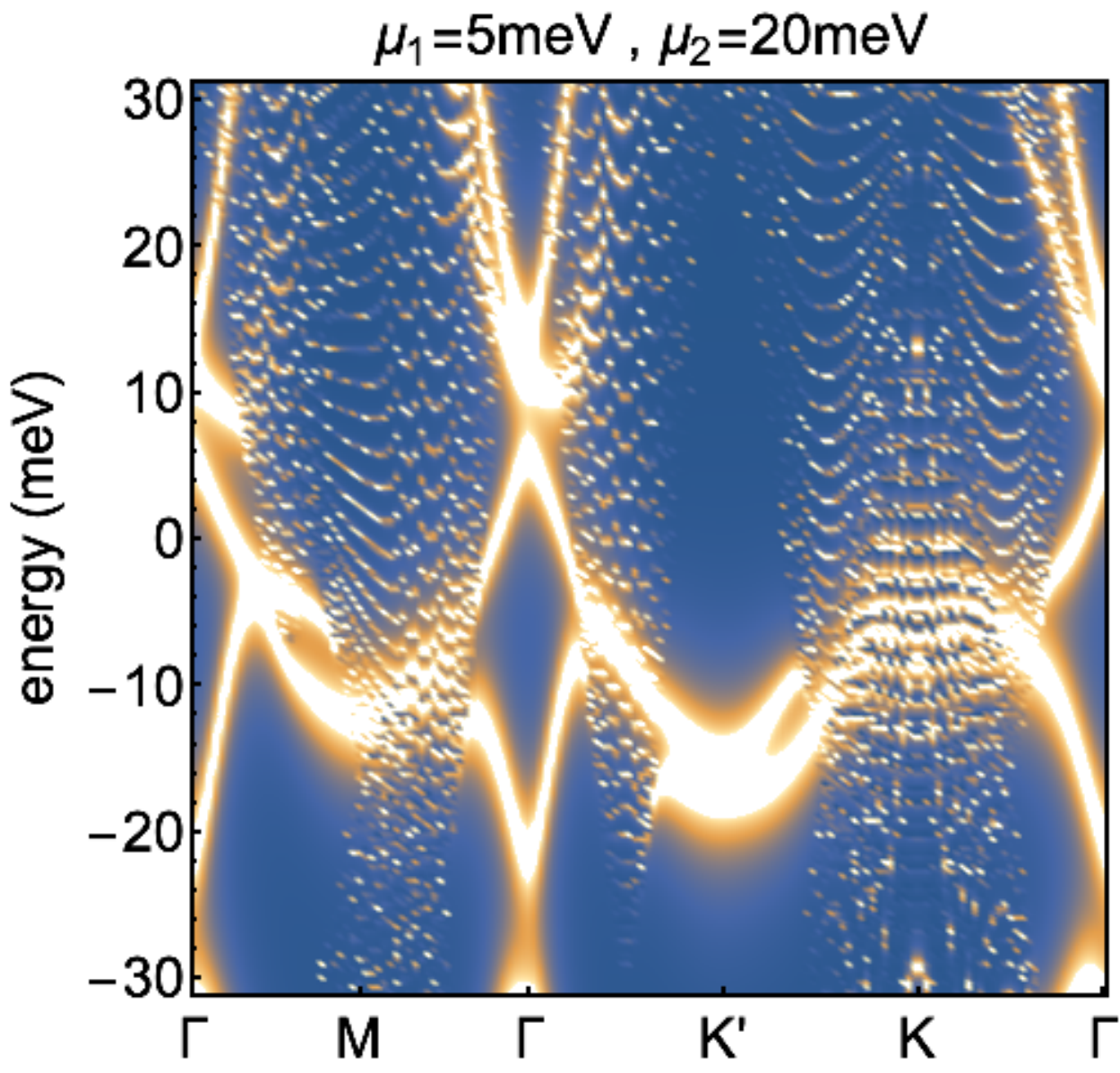} &
\includegraphics[width=2.in]{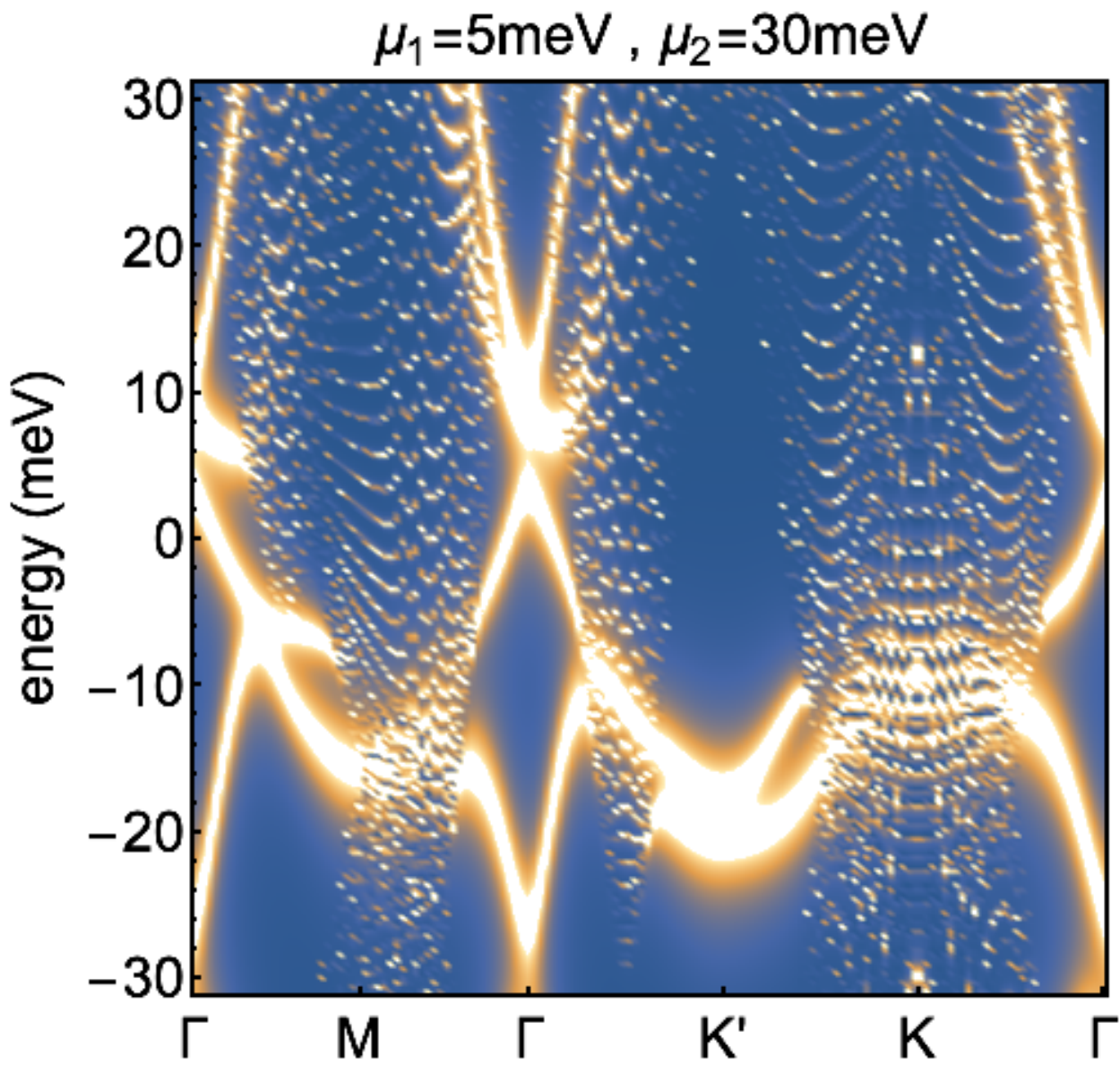} &
\includegraphics[width=0.5in]{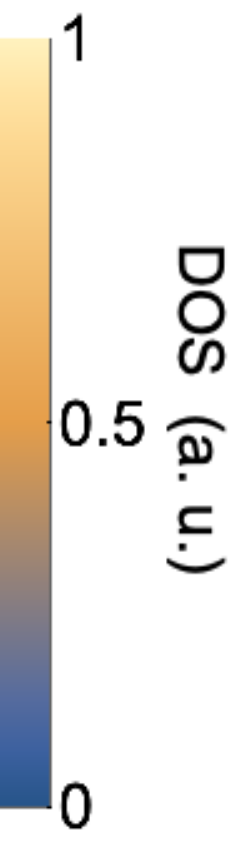} 
\end{tabular}
\caption{
DOS of the twisted monolayer on top of graphite in the presence of an applied gate at the interface.}
\lb{gate}
\end{figure}
The changes due to the gate potential are rather small, although noticeable. The small effect of the gate is probably due to level repulsion between the low energy band and the continuum coming from the semi-infinite stack.

\section{Lattice Relaxation}

The idea of the lattice relaxation is discussed in great detail in Refs.~\cite{WG19si,GW19si}, and the results presented here are a based on the work there. We use classical potential models as implemented in the LAMMPS code \cite{plimpton_fast_1995} to relax the atomic positions. The potential models used here are the AIREBO-M\cite{oconnor_airebo-m:_2015si} for inlayer interactions, overlaid by the Kolmogorov-Crespi potential \cite{kolmogorov_registry-dependent_2005si} for the long-range interactions between the layers.

\begin{figure}[ht!]
\includegraphics[width=5in]{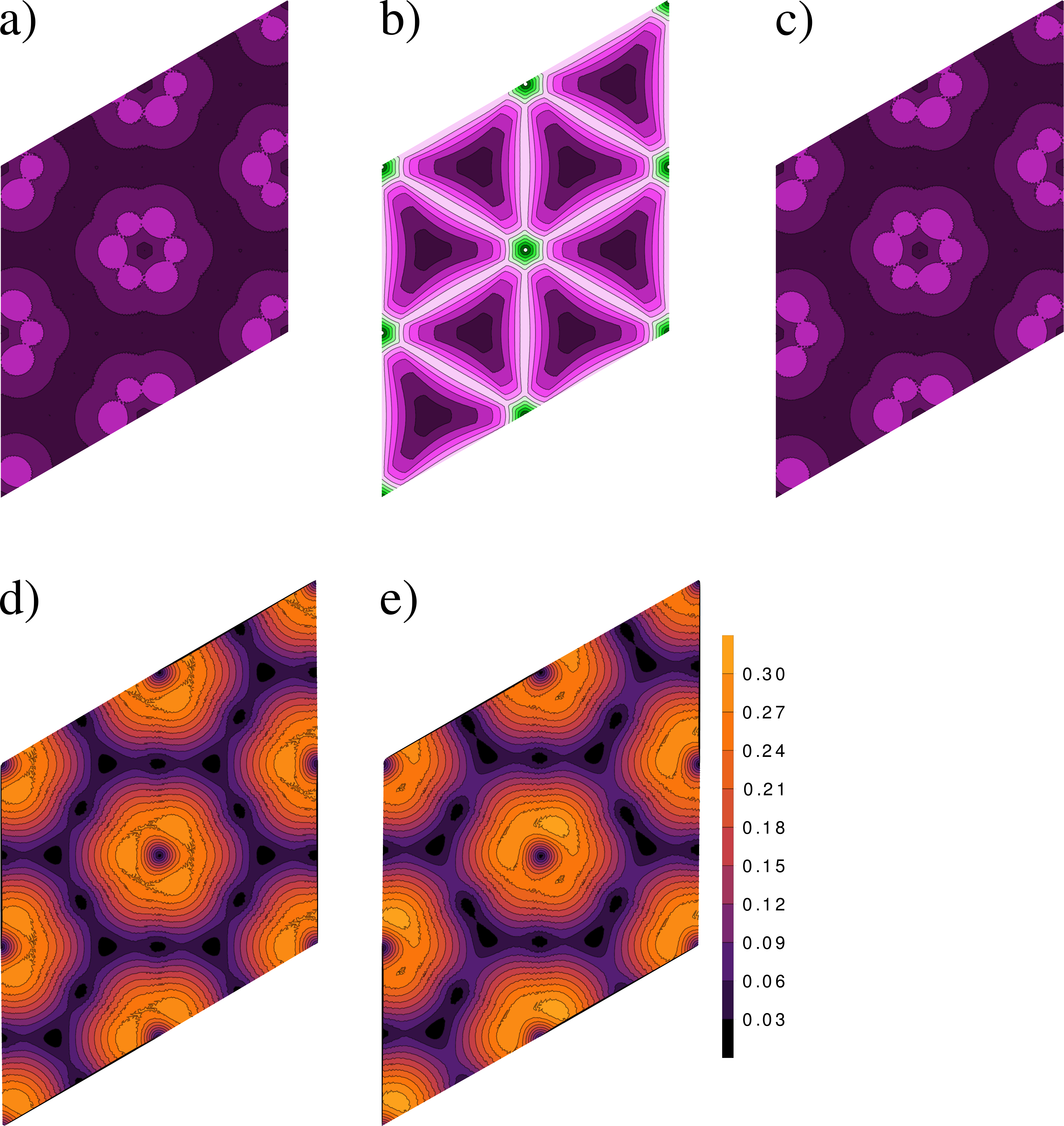}
\caption{Relaxation of twisted graphite on graphite. a) alignment of the top layer of the bottom  graphite stack; b) alignment of the top layer of stack 1 with the lower layer of stack 2; c)  alignment of the bottom layer of the top  graphite stack; c) absolute value of the displacement of the top layer of stack 1; d) similar for the bottom layer of stack 2.}\label{fig:graphgraph}
\end{figure}

We have modelled the atomic positions for the last two problems only using this approach, since the first one is not easily attacked by such a simulations due to the lack of periodicity. We show the results for the single graphene layer in Fig.~5 of the main text. We model the graphite as an $ABA$ stack with the lowest two $AB$ layers completely fixed, at the graphite spacing. The top $A$ layer and the rotated graphene layer are left completely free.  
We have performed a similar calculation for graphite on graphite, where each stack is modelled as a movable $A$ layer on a fixed $B$ layer, and we assume all layers are spaced at the graphite layer spacing, which fortunately is very close to the equilibrium graphene-graphene spacing. As we can see in Fig.~\ref{fig:graphgraph} the surface reconstruction is very similar to that for graphene on graphite: The alignment between the two movable layers seems very similar. This means that the effect of relaxation should be rather universal as well, since the hopping parameters at the twist interfaces are expected to be similar.

\input{SUPPLEMENTARY.bbl}\end{document}

%% file: twisted_graphite.bbl
%

%% file: twisted_graphite.bbl
\begin{thebibliography}{63}
\bibitem [{not()}]{note2}%
  \BibitemOpen
  \href@noop {} {}\bibinfo {note} {The parameter $\gamma_2$ will also give a
  broadening to the low energy bands in the geometry discussed in\cite{KKTV19},
  as second nearest neighbor layers are in registry.}\BibitemShut {Stop}%
\end{thebibliography}
